\documentclass[]{jfm}
\usepackage{bm} \usepackage{amssymb} \usepackage{natbib}
\usepackage{graphicx} \usepackage{color}
\usepackage{amsmath}
\usepackage{amsfonts}
\usepackage{bm}
\usepackage{placeins}
\newcommand{\be}{\begin{equation}}
\newcommand{\ee}{\end{equation}}
\usepackage{soul}
\usepackage{color}

\usepackage{bm}
\usepackage{amssymb}
\usepackage{natbib}
\usepackage{graphicx}
\usepackage{color}
\usepackage{amsmath}
\usepackage{amsmath}
\usepackage{amsfonts}
\usepackage{bm}
\usepackage{placeins}
\usepackage{graphicx}
\usepackage{booktabs}
\usepackage{array}
\usepackage{hyperref}

\usepackage{bm}
\usepackage{bbm}
\usepackage{times}

\usepackage{soul}
\usepackage{color}

\newcommand{\D}{\mathrm{d}}

\newcommand{\bea}{\begin{eqnarray}}
\newcommand{\eea}{\end{eqnarray}}

\begin{document}

\title[Plastic flow of foams and emulsions in a channel]{Plastic flow of foams and emulsions in a channel: experiments and simulations}

\author[B. Dollet, A. Scagliarini and M. Sbragaglia]{B. \ns D\ls O\ls L\ls L\ls E\ls T\ls $^1$, A.\ns S\ls C\ls A\ls G\ls L\ls I\ls A\ls R\ls I\ls N\ls I $^{2}$ and  M.\ns S\ls B\ls R\ls A\ls G\ls A\ls G\ls L\ls I\ls A\ls $^{2}$}

\affiliation{$^{1}$ Institut de Physique de Rennes, UMR 6251 CNRS/Universit\'e Rennes 1, Campus Beaulieu, B\^atiment 11A, 35042 Rennes Cedex, France \\ $^{2}$ Department of Physics and INFN, University of Rome Tor Vergata, \\ Via della Ricerca Scientifica 1, 00133 Rome, Italy }

\date{\today}
\maketitle

\begin{abstract}
In order to understand the flow profiles of complex fluids, a crucial issue concerns the emergence of spatial correlations among plastic rearrangements exhibiting cooperativity flow behaviour at the macroscopic level. In this paper, the rate of plastic events in a Poiseuille flow is experimentally measured on a confined foam in a Hele-Shaw geometry. The correlation with independently measured velocity profiles is quantified. To go beyond a limitation of the experiments, namely the presence of wall friction which complicates the relation between shear stress and shear rate, we compare the experiments with simulations of emulsion droplets based on the lattice-Boltzmann method, which are performed both with, and without, wall friction. Our results indicate a correlation between the localisation length of the velocity profiles and the localisation length of the number of plastic events. Finally, unprecedented results on the distribution of the orientation of plastic events show that there is a non-trivial correlation with the underlying local shear strain. These features, not previously reported for a confined foam, lend further support to the idea that cooperativity mechanisms, originally invoked for concentrated emulsions \citep{Goyon08}, have parallels in the behaviour of other soft-glassy materials.
\end{abstract}


\section{Introduction}\label{sec:intro}

Foams and emulsions are dispersions of a fluid phase in a liquid phase, stabilised by surfactants. The dispersed phase is constituted of gas bubbles in foams, and liquid droplets in emulsions. These discrete objects are packed together and jammed, which makes foams and emulsions complex fluids: they exhibit a yield stress $\sigma_Y$ below which they do not flow, but deform elastically. Above yield stress, they flow like rheothinning fluids. Rheometric measurements in a Couette cell or in cone--plate geometry have shown that the shear stress $\sigma$ and the shear rate $\dot{\gamma}$ obey an empirical Herschel--Bulkley law: $\sigma = \sigma_Y + A\dot{\gamma}^n$, with $A$ the plastic viscosity and $n$ an exponent generally lower than 1, and often close to 0.5 \citep{Princen89,Marze08,Denkov09}, with some dependence on the surfactants used \citep{Denkov09}.

The aforementioned measurements did not give access to the microstructure under flow, and other techniques have been developed to visualise it. In emulsions, confocal microscopy on systems of matched optical index have recently enabled to measure the local structure \citep{Jorjadze13} and the velocity field \citep{Goyon08,Goyon10,Mansard14}. The latter could also be measured using magnetic resonance imaging \citep{Ovarlez08}. In foams, index matching is not possible, and the route has been to devise bidimensional (2D) experiments, on either bubble rafts at the surface of a pool of soap solution, with or without a confining top plate \citep{Lauridsen04,Dollet05,Wang06,Katgert08,Katgert10}, or on bubble monolayers confined between two plates in a Hele-Shaw cell \citep{Debregeas01}.

Among many interesting features such as shear banding (see e.g. \citet{Schall10} for a review), these studies have called the Herschel--Bulkley law found in rheometry into question. Among possible flow configurations, the Poiseuille flow in a straight channel is particularly interesting, since this geometry enforces a linear variation across the channel of the shear stress, which vanishes at the centre and reaches its maximum at the side walls. Together with an evaluation of the shear rate from the measured velocity profile, it gives access to the relation $\sigma(\dot{\gamma})$ at the local scale. In particular, \citet{Goyon08,Goyon10} have measured this relation in a series of experiments on emulsions, and they have shown that it did not collapse on a single Herschel--Bulkley law. This deviation from a single flow curve was ascribed to wall effects, more precisely to a nonlocal influence of plastic events happening in the vicinity of the boundaries. The velocity profiles were convincingly fitted by a fluidity model \citep{Goyon08,Goyon10}. This model, based on a kinetic theory approach \citep{Bocquet09}, predicted that the fluidity, defined as $f = \dot{\gamma}/\sigma$, is proportional to the rate of plastic events and follows a nonlocal diffusion equation when it deviates from its bulk value. The range of influence $\xi$ appearing in this equation, called the {\it spatial cooperativity}, was shown to be of the order of a few times (typically, five) the size of the elementary microstructural constituent (the drop in the case of emulsions) \citep{Goyon08,Goyon10,Geraud13}. This picture was later applied to other complex fluids, such as Carbopol gels \citep{Geraud13}, granular media \citep{Amon12,Kamrin12}, and foams in a 2D cylindrical Couette geometry \citep{Katgert10}. The fluidity model agrees with existing experiments, and provides a convenient framework to rationalise the flow of complex fluids. However, at least two points remain unclear and deserve further investigation. The first is the boundary condition at solid walls for fluidity. As a matter of fact, most experimentalists have set it as a free fit parameter, which certainly improves the agreement between the measurements and the predictions from the fluidity model, but does not provide any insight on the role of the walls. Only recently, \citet{Mansard14} explored the role surface boundary conditions for the flow of a dense emulsion. They show that both slippage and wall fluidisation depend non-monotonously on the roughness, a behaviour that has been interpreted with a simple model invoking the building of a stratified layer and the activation of plastic events by the surface roughness. These results are interesting and call for further verification in terms of numerical simulations \citep{EPL13,Sbragaglia12} and other complex fluids \citep{Katgert10}. Second, the fluidity parameter $f$ has not been yet convincingly related to an independent measure of the local density of plastic events. In experiments, only indirect indications of such a relation have been proposed, based on the correlations of the fluctuations of the shear rate \citep{Jop12}. Using numerical simulations based on the bubble model \citep{Durian97}, \citet{Mansard13} were able to measure independently the fluidity and the density of plastic events, but they show that the two quantities are not proportional; more precisely, the rearrangement rate was found to be a sublinear power (with an exponent 0.4) of the fluidity.

Actually, fluidity models offer a potential explanation for the deviation from a unique relation between stress and strain rate, but they are not the only ones. Another approach has been to develop elasto-viscoplastic models (see e.g. \citet{Cheddadi12} for a review of them) which, in essence, supplement the viscoplastic Herschel--Bulkley rheology by a description of elasticity. These models are local, but since they treat elastic deformation as an independent variable, they also predict deviations from a single Herschel--Bulkley relation. They have been compared with experiments in Couette flows \citep{Cheddadi12}, but not for Poiseuille flows.

All these theoretical approaches rely crucially on the modelling of plastic events, and how they affect the elastic stress and the flow. However, although this connection between elasticity, plasticity and flow has been studied in foam flows in complex geometries \citep{Dollet07,Dollet10,Cheddadi11}, there is no existing experimental measurement of the rate of plastic events in a Poiseuille flow. 2D foams are particularly well suited for such a study, because elementary plastic events (so-called T1 events) are well characterised by the neighbour swapping of four bubbles (Figs.~\ref{Fig:sketch_T1} and \ref{Fig:sketch_T1_sim}) and are accessible by image analysis, more easily that in other soft glassy materials.

In this paper, we provide the first experimental measurements of the rate of plastic events in a Poiseuille flow, on a confined foam in a Hele-Shaw geometry. We show that it is closely related to the independently measured velocity profiles, and that there is still a non-vanishing plastic activity towards the centre of the channel. The study of the spatial distribution in the number of plastic events and the simultaneous analysis of the velocity profiles allows to bridge between the details of the irreversible plastic rearrangements and the corresponding cooperativity flow behaviour at the macroscopic level \citep{Goyon08,Goyon10,Geraud13}. We choose to explore this connection by looking at the relationship between the localisation length of the velocity profiles and the localisation length of the number of plastic events. In our experiments, because of wall friction, there is no simple relation between shear stress and shear rate. Therefore, we compare the experiments with simulations of emulsion droplets based on the lattice-Boltzmann method \citep{Sbragaglia12}, which are performed both with, and without, wall friction. Numerical simulations also offer the possibility to test the robustness of some of the experimental findings versus a change in the viscous ratio $\chi$ between the dispersed phase and the continuous phase, this being set to $\chi=1$ in all the numerical simulations, whereas $\chi \approx 10^{-2}$ in foams; in that sense, the simulations look closer to emulsions. The numerical model possesses two advantages that are rarely present together. From one side, it gives a realistic structure of the emulsion droplets, like for example the Surface Evolver method \citep{Surface1,Surface2,Surface3}; at the same time, due to the built-in properties, the model gives direct access to equilibrium and out-of-equilibrium stresses \citep{Sbragaglia12}, including elastic and the viscous contributions. In contrast to other mesoscopic models, such as Durian's bubble model \citep{Durian97}, our model naturally incorporates the dissipative mechanisms and the interfacial stresses.

The paper is organised as follows. In Sec.~\ref{sec:experiments}, we describe the experimental set-up along with the tools of image analysis for characterisation of the plastic events. In Sec~\ref{sec:LBM}, and supplementary material presented in Appendices \ref{AppendixA} and \ref{AppendixB}, we review our computational model based on the lattice Boltzmann models (LBM). The review of the computational model will be accompanied by further benchmark tests on the capability of the model to include crucial properties as disjoining pressure and friction. Results and discussions will be the subject of Sec.~\ref{sec:ResultsandDiscussion}. The experimentally measured velocity profiles (Sec.~\ref{Subsec:experimental_velocity_profiles}) will be compared with local linear and nonlinear models (Sec.~\ref{Subsec:local_model} and Appendix \ref{AppendixC}). Results of numerical simulations and comparisons with the fluidity model \citep{Goyon08,Bocquet09} will be the subject of Sec.~\ref{sec:numexp}. In Sec.~\ref{Subsubsec:plasticflow}, we compare the localisations of the velocity profiles and of the rate of plastic events. In Sec.~\ref{ref:orientation}, we will finally report details on the orientation of the plastic rearrangements in the flowing material. Conclusions will follow in Sec.~\ref{sec:conclusion}.

\section{Experimental methods}\label{sec:experiments}

\subsection{Setup}\label{subsec:MATMETHODS}

\begin{figure}
\begin{center}
\includegraphics[width=12cm,keepaspectratio]{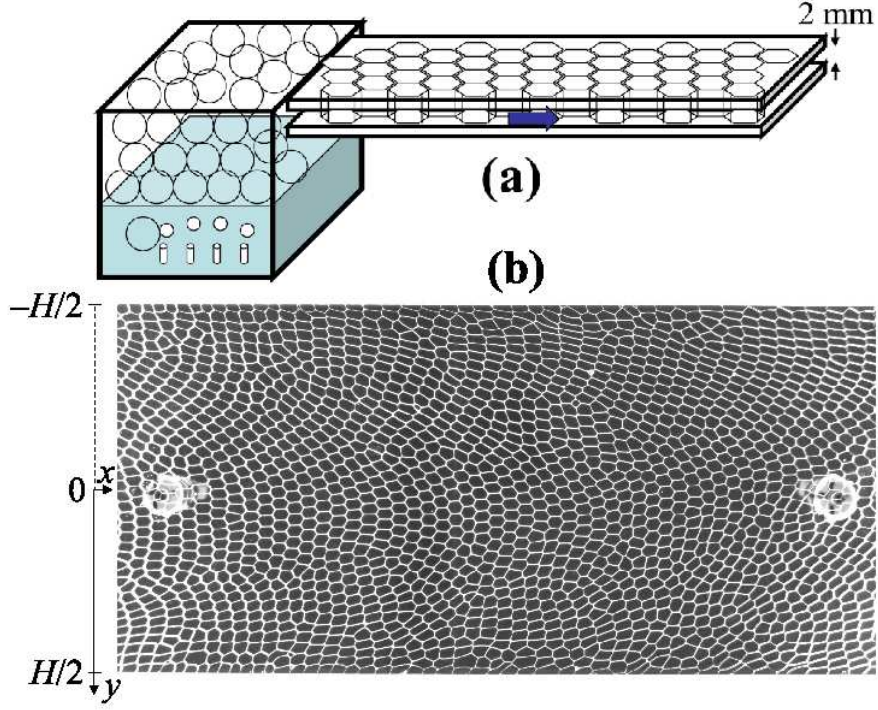}
\caption{(a) Sketch of the side view of the setup. (b) Snapshot of an experiment. The distance between the two side walls is $H=106.6$~mm. The average bubble size is 12.3~mm$^2$ and the liquid fraction is $\phi_l=4.8\%$. The two spots at the left and right of the image are the points between which the pressure drop is measured.}
\label{Fig:setup_Benji}
\end{center}
\end{figure}

We have adapted the setup described in \citet{Dollet10}. The foam flows in a Hele-Shaw cell, made of two horizontal glass plates of length 170~cm and width 32~cm, separated by a gap $h=2$~mm thin enough that the foam is confined as a bubble monolayer (Fig.~\ref{Fig:setup_Benji}a). Two plastic plates of thickness 2~mm are inserted aside the Hele-Shaw cell, so that the width $H$ of the channel is reduced to 10.66~cm (Fig.~\ref{Fig:setup_Benji}b). These plates have a negligible roughness compared to the bubble size. The channel is connected upstream to a vertical chamber (Fig.~\ref{Fig:setup_Benji}a) in which a soap solution is fed at a prescribed flow rate $Q_l$ thanks to a syringe pump (PHD2000, Harvard Apparatus). Nitrogen is continuously blown through injectors at the bottom of this chamber, producing rather monodisperse bubbles (Fig.~\ref{Fig:setup_Benji}b). The flow rate in each injector is independently controlled with an electronic flow-rate controller (Brooks). We identify the liquid fraction $\phi_l$ as the ratio of the liquid flow rate to the total flow rate: $\phi_l = Q_l/(Q_g + Q_l)$, with $Q_g$ the gas flow rate. The resulting foam accumulates on top of the chamber, over a vertical distance where it drains, then is pushed through the channel. The transit time through the whole channel is less than 10 minutes in all experiments; we do not observe significant change of bubble size during this time, hence coarsening is negligible. The soap solution is a mixture of sodium lauryl-dioxyethylene sulfate (SLES), cocoamidopropyl betaine (CAPB) and myristic acid (MAc), following the protocol described in \citet{Golemanov08}: we prepare a concentrated solution of 6.6\% wt of SLES and 3.4\% of CAPB in ultra-pure water, we dissolve 0.4\% wt of MAc by continuously stirring and heating at 60$^\circ$C for about one hour, and we dilute 20 times in ultra-pure water. The solution has a surface tension $\Gamma = 22.4$~mN/m. The contraction region is lit by a circular neon tube, giving an isotropic and nearly homogeneous illumination over a diameter of about 20~cm. Movies of the foam flow are recorded with a CCD camera at a frame rate of 8 frames per second, with an exposure time of 8~ms. The movies are constituted of 1000~images of $1312\times 672$~pixels. The pressure drop is measured across the observation zone, by a water--water differential manometer connected to two points of the channel (Fig.~\ref{Fig:setup_Benji}b) through tubes full of water. We have performed five different experiments, a summary of which parameters is provided in Tab.~\ref{Tab:summary_exp}.

\begin{table}
  \centering
  \begin{tabular}{|c|c|c|c|c|c|c|c|c|c|c|}
  \hline
  flow rate & $\phi_l$ & bubble & $v_0$ & $L_v$ & $\alpha_s$ & $\theta_d$ & $\theta_a$ & $\mathrm{d}P/\mathrm{d}x$ & $\mathrm{d}\sigma_{xy}/\mathrm{d}y$ & symbol \\
  (ml/min) & (\%) & area (mm$^2$) & (mm/s) & (mm) &  &  (rad) & (rad) & (kPa/m) & (kPa/m) & \\
  \hline
  27.5 & 4.8 & $13.0\pm 2.4$ & 2.3 & 3.3 & 0.62 & $0.54\pm 0.14$ & $-0.79\pm 0.24$ & 1.08 & 0.50 & $\circ$ \\
  52.5 & 4.8 & $12.6\pm 2.3$ & 4.3 & 3.9 & 0.43 & $0.46\pm 0.16$ & $-0.67\pm 0.33$ & 1.08 & 0.68 & $\square$ \\
  102.5 & 4.8 & $12.1\pm 2.1$ & 8.5 & 8.7 & 0.34 & $0.46\pm 0.19$ & $-0.75\pm 0.46$ & 2.23 & 0.51 & $\lozenge$ \\
  152.5 & 4.8 & $12.3\pm 1.7$ & 12.4 & 6.8 & 0.31 & $0.46\pm 0.17$ & $-0.73\pm 0.41$ & 2.27 & 0.91 & $\vartriangle$ \\
  160.2 & 16.9 & $15.1\pm 1.9$ & 12.9 & 4.6 & 0.90 & $0.73\pm 0.22$ & $-0.86\pm 0.21$ & 0.91 & 0.087 & $\triangledown$ \\
  \hline
\end{tabular}
  \caption{Summary of the main characteristics of the experiments presented in this paper, with the respective symbols used in the figures. The parameters $v_0$, $\alpha_s$ and $L_v$ come from the fit of the velocity profiles by the formula (\ref{Eq:fit_velocity}). The angles $\theta_d$ and $\theta_a$ are the orientations of plastic events (Figs. \ref{Fig:sketch_T1} and \ref{Fig:sketch_T1_sim}). The quantity $\D P/\D x$ is the pressure drop along the channel, and $\D\sigma/\D y$ is the gradient of shear elastic stress across the channel.}
  \label{Tab:summary_exp}
\end{table}

\subsection{Image analysis}

\begin{figure}
\begin{center}
\includegraphics[width=12cm,keepaspectratio]{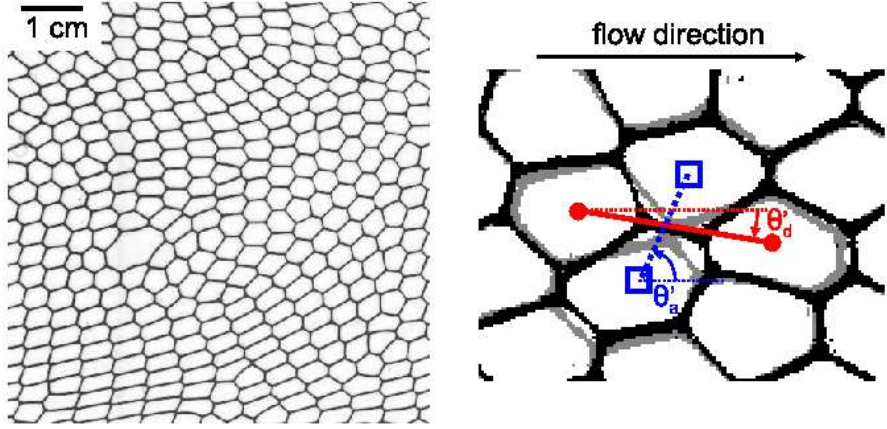}
\caption{Left Panel: A snapshot of the bubbles in the foam flowing from left to right. Right Panel: Sketch of a plastic event. By following the displacements of the bubbles between subsequent images, we are able to determine the features of a T1 rearrangement. In gray (black) we report the bubble edges just before (after) the T1. With the solid (dashed) line, we report the link between the centres of the two bubbles that lose (come into) contact during the T1. From the analysis of the links, we are able to determine the angles associated with the links that disappear (d) or appear (a) in the T1 rearrangement.}
\label{Fig:sketch_T1}
\end{center}
\end{figure}

\begin{figure}
\begin{center}
\includegraphics[scale=0.5]{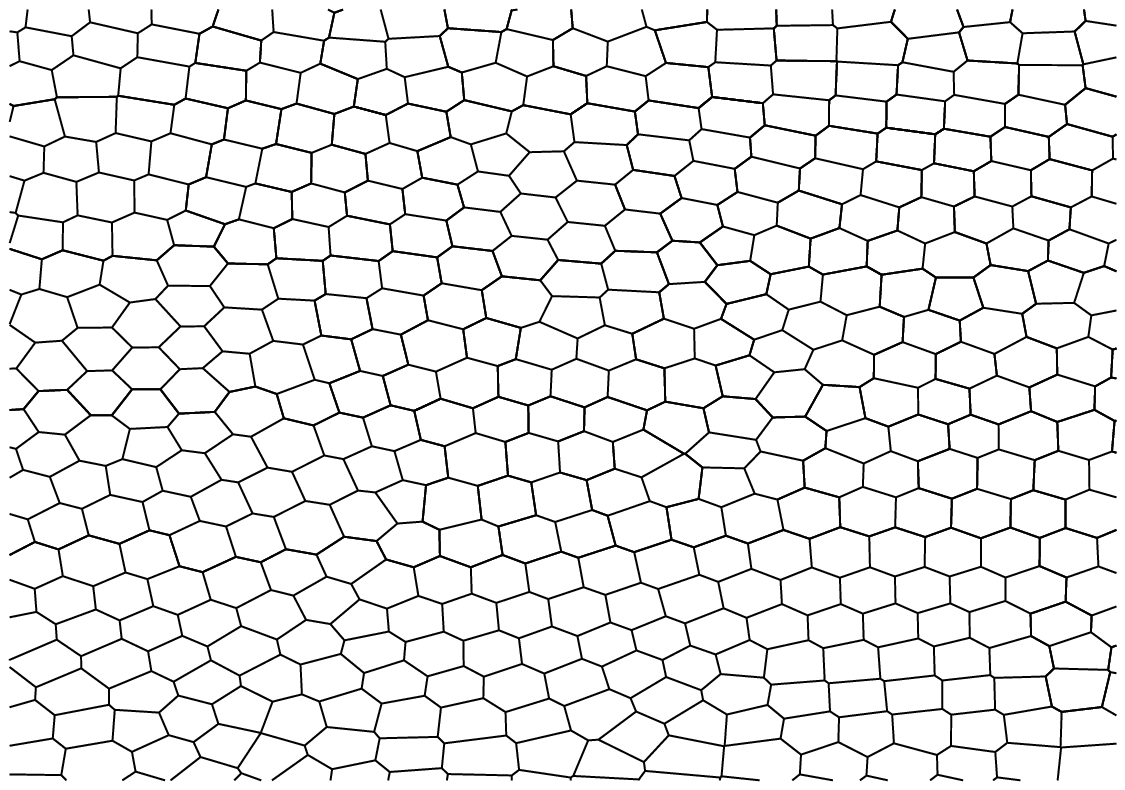}
\includegraphics[scale=0.5]{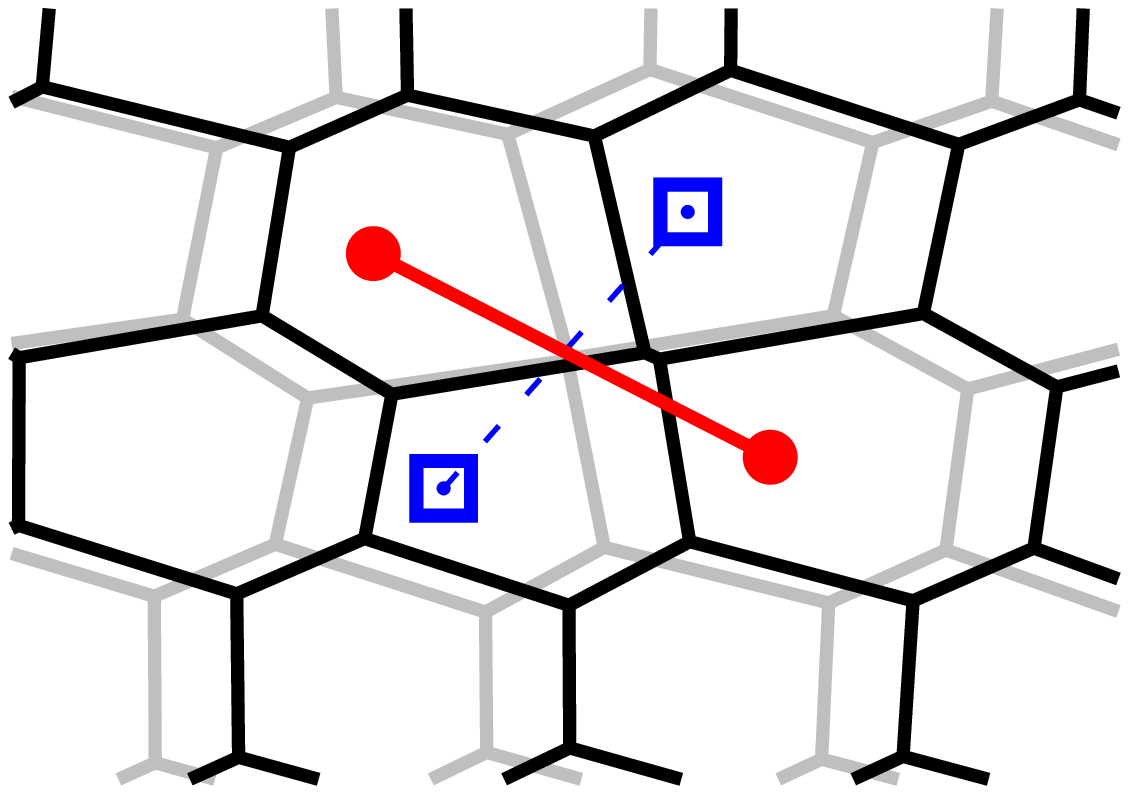}
\caption{Left Panel: A snapshot of the droplets (identified by their corresponding Voronoi cells) in a concentrated emulsion, flowing from left to right, obtained in numerical simulations based on the lattice Boltzmann models. Right Panel: sketch of a T1 plastic event from the simulations. To systematically analyze plastic events, we perform a Voronoi tessellation from the centres of mass of the droplets. Following the Voronoi tessellation in time, we are able to identify T1 events and associated disappearing (red solid line) and appearing (blue dashed line) links. In gray (black) we indicate the Voronoi cells soon before (after) a T1 event. The numerical results will be compared with the experimental results (see also Fig.~\ref{Fig:sketch_T1}).}
\label{Fig:sketch_T1_sim}
\end{center}
\end{figure}

To extract the relevant rheological information from the movies, we follow a home-made procedure very similar to that presented in \citet{Dollet07} and \citet{Dollet10}; we report to these papers for full details. The velocity field is obtained after averaging of all the displacements of all individual bubbles between consecutive frames (about $3\times 10^6$ in total). Averaging is performed along 53 lanes aligned with the flow direction. The T1s are tracked as described in \citet{Dollet07}. For the four bubbles concerned by a T1, we denote $\mathbf{r}_d$ ($\mathbf{r}_a$) the vector linking the centres of the two bubbles that lose (come into) contact, $\theta'_d$ and $\theta'_a$ the angle of these vectors with respect to the flow direction, that we can restrict to the interval $[-\pi/2,\pi/2]$ because the orientations of $\mathbf{r}_d$ and $\mathbf{r}_a$  are irrelevant (Fig.~\ref{Fig:sketch_T1} and Fig.~\ref{Fig:sketch_T1_sim}), and $\mathbf{x}_d$ ($\mathbf{x}_a$) the position of the midpoint of the centres of the two bubbles that lose (come into) contact. In our program, the detection of appearing and disappearing contacts is first run independently. As a second step, to identify a T1 and minimise artefacts, we decide that a pair of an appearing and a disappearing contact constitute a single T1 if (i) they are on the same image or if the appearing contact happens in the image next to the disappearing one; the latter condition is necessary, because it happens that transient fourfold vertices are erroneously recognised as artificial small bubbles; (ii) the positions $\mathbf{x}_d$ and $\mathbf{x}_a$ are closer than a critical distance (that we choose to be of the order of the bubble size, to separate from T1s occurring in the neighborhood); (iii) $|\theta'_d - \theta'_a|$ is larger than a critical angle (we choose $\pi/4$), this condition being necessary because of the apparition of the aforementioned spurious bubbles. By visual inspection on 30 images, we estimate that this procedure leads to an uncertainty of no more than 5\% on the number $N_{T1}$ of T1s. We then define the quantity $(\mathbf{x}_d + \mathbf{x}_a)/2$ as the position of a T1, and we ascribe this information to the box where this position belongs. We thus compute the scalar field of the frequency of T1s per unit time and area:
$$
f_{T1} = \frac{N_{T1}}{2A_{\mathrm{box}} t_{\mathrm{movie}}},
$$
where $A_{\mathrm{box}}$ is the area of a box and $t_{\mathrm{movie}}$ the duration of a movie. Our T1 detection has two major advantages: (i) it is directly based on the topological rearrangements, contrary to indirect characterisations based on velocity correlations; (ii) it yields an unprecedented statistics, up to $2.5\times 10^4$ individual T1s, which enables to average over the same lanes as for the velocity and to perform quantitative analysis.

Finally, a specific advantage of 2D foams is the possibility to measure directly the elastic stress from image analysis. Neglecting the curvature of the bubble edges, the 2D elastic stress tensor writes \citep{Batchelor70,Cantat13}:
\begin{equation}\label{Eq:elastic_stress_tensor}
\boldsymbol\sigma_{\mathrm{2D}} = \lambda\rho_{\boldsymbol\ell} \left\langle
\frac{\boldsymbol\ell\otimes\boldsymbol\ell}{\ell} \right\rangle ,
\end{equation}
with $\lambda \simeq 2\gamma h$ the line tension, $\rho_{\boldsymbol\ell}$ the areal density of bubble edges, and where the average is computed over all bubble edges $\boldsymbol\ell$. Assuming that the films are invariant between the top and bottom plates, the 3D elastic stress equals: $\boldsymbol\sigma = \boldsymbol\sigma_{\mathrm{2D}}/h$.


\section{Numerical method}\label{sec:LBM}

\begin{figure}
\begin{center}
\includegraphics[width=8.0cm,keepaspectratio]{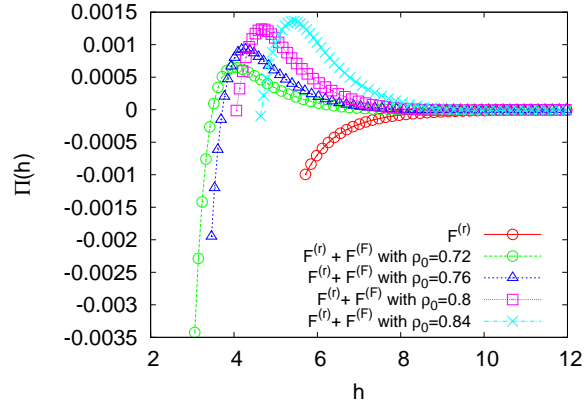}
\caption{This figure shows the emergence of the disjoining pressure in the lattice Boltzmann model (see Eq.~\eqref{LB} with phase separating interactions obtained with a repulsive ($r$) force  (see Eq.~\eqref{Phase}) supplemented with competing interactions (see Eq.~\eqref{NNandNNN}) whose role is to provide a mechanism for {\it frustration} ($F$). The use of phase separating interaction is associated with a negative disjoining pressure. Competing interactions stabilise thin films with the emergence of a positive disjoining pressure, the latter tunable with the parameter $\rho_{0}$ in Eqs.~\eqref{Phase} and \eqref{NNandNNN}. Further details can be found in \citet{Sbragaglia12}. \label{fig:disjoining}}
\end{center}
\end{figure}

For the numerical simulations, we adopt a dynamic rheological model based on the lattice Boltzmann method (LBM) \citep{Benzi92,Chen98,Aidun}.  Historically, the main successful applications of LBM in the context of computational fluid dynamics pertain to the weakly compressible Navier--Stokes equations \citep{Benzi92,Chen98} and models associated with more complex flows involving phase transition/separation \citep{SC1,SC2,CHEM09}. In particular, we will make use of a computational model for non-ideal binary fluids, which combines a positive surface tension, promoting the formation of diffuse interfaces, with a positive disjoining pressure, inhibiting droplet (or bubble) coalescence. The model has already been described in several previous works \citep{CHEM09,EPL10,EPL13}.  In this section we review the method and highlight its essential supramolecular features. The mesoscopic kinetic model considers two fluids $A$ and $B$, each described by a {\it discrete} kinetic distribution function $f_{\zeta i}(\mathbf{r},\mathbf{c}_i,t)$, measuring the probability of finding a particle of fluid $\zeta =A,B$ at position $\mathbf{r}$ and time $t$, with discrete velocity $\mathbf{c}_i$. In other words, the mesoscale particle represents all molecules contained in a unit cell of the lattice. The distribution functions evolve in time under the effect of free-streaming and local two-body collisions, described, for both fluids ($\zeta=A,B$), by a relaxation towards a local equilibrium ($f_{\zeta i}^{(eq)}$) with a characteristic time $\tau_{LB}$:
\begin{equation}
\label{LB}
f_{\zeta i}(\mathbf{r}+\mathbf{c}_i,\mathbf{c}_i,t+1) -f_{\zeta i}(\mathbf{r},\mathbf{c}_i,t)  = -\frac{1}{\tau_{LB}} \left(f_{\zeta i}-f_{\zeta i}^{(eq)} \right)(\mathbf{r},\mathbf{c}_i,t)+F^{(tot)}_{\zeta i}(\mathbf{r},\mathbf{c}_i,t).
\end{equation}
Local equilibria are given by a low Mach number expansion of the Maxwellian distribution, namely:
\be
f_{\zeta  i}^{(eq)}=w_i \rho_{\zeta} \left[1+\frac{\mathbf{v} \cdot \mathbf{c}_i}{c_s^2}+\frac{\mathbf{v}\mathbf{v}:(\mathbf{c}_i\mathbf{c}_i-c_s^2 \bar{\bar{1}})}{2 c_s^4} \right]
\ee
with $w_i$ a set of weights chosen in such a way to maximise the algebraic degree of precision in the computation of the hydrodynamic fields, while $c_s=1/\sqrt{3}$ is a characteristic velocity (a constant in the model). Our lattice scheme features nine discrete velocities \citep{Shan06}, whose details and associated weights are reported in Tab.~\ref{TABLE} in Appendix A. Coarse grained hydrodynamical densities are defined for both species $\rho_{\zeta }=\sum_i f_{\zeta i}$ as well as a global momentum for the whole binary mixture $\mathbf{j}=\rho \mathbf{v}=\sum_{\zeta , i} f_{\zeta i} \mathbf{c}_i$, with $\rho=\sum_{\zeta} \rho_{\zeta}$. Non-ideal forces ($\mathbf{F}_{\zeta}$)  and a body force term ($\mathbf{F}_{b}$) are introduced with the source term $\mathbf{F}^{(tot)}_{\zeta}$ in Eq.~\eqref{LB}. The non-ideal forces include a variety of interparticle forces, $\mathbf{F}_{\zeta}=\mathbf{F}^{(r)}_\zeta+\mathbf{F}^{(F)}_\zeta$. First, a repulsive ($r$) force with strength parameter ${\cal G}_{AB}$ between the two fluids
\begin{equation}\label{Phase}
\mathbf{F}^{(r)}_\zeta (\mathbf{r})=-\frac{{\cal G}_{AB}}{\rho_0^2} \rho_{\zeta }(\mathbf{r}) \sum_{i=1-8, \zeta ' \neq \zeta } w_i \rho_{\zeta '}(\mathbf{r}+\mathbf{c}_i)\mathbf{c}_i
\end{equation}
is responsible for phase separation \citep{CHEM09}. The parameter $\rho_0$ is a characteristic normalisation parameter, used as a free parameter in the model. The ``short'' range interaction in Eq.~\eqref{Phase} is extended up to energy shells $|\mathbf{c}_i|^2=2$ (lattice links have been normalised to a characteristic lattice velocity). Furthermore, both fluids are also subject to competing interactions whose role is to provide a mechanism for {\it frustration} ($F$) for phase separation. In particular, we model short range (nearest neighbor, NN) self-attraction, controlled by strength parameters ${\cal G}_{AA,1} <0$, ${\cal G}_{BB,1} <0$), and ``long-range'' (next to nearest neighbor, NNN) self-repulsion, governed by strength parameters ${\cal G}_{AA,2} >0$, ${\cal G}_{BB,2} >0$)
\begin{equation}\label{NNandNNN}
\mathbf{F}^{(F)}_\zeta (\mathbf{r})=-{\cal G}_{\zeta \zeta ,1} \psi_{\zeta }(\mathbf{r}) \sum_{i=1-8} w_i \psi_{\zeta }(\mathbf{r}+\mathbf{c}_i)\mathbf{c}_i -{\cal G}_{\zeta \zeta ,2} \psi_{\zeta }(\mathbf{r}) \sum_{i=1-24} p_i \psi_{\zeta }(\mathbf{r}+\mathbf{c}_i)\mathbf{c}_i
\end{equation}
with $\psi_{\zeta }(\mathbf{r})=\psi_{\zeta }[\rho(\mathbf{r})]$ a suitable pseudo-potential function. The pseudo-potential $\psi_\zeta(\rho_\zeta)$ is taken in the form originally suggested by \citet{SC1,SC2}
\begin{equation}
\label{PSI}
\psi_{\zeta}[\rho_{\zeta}(\mathbf{r})]= \rho_{0} [1-\mathrm{e}^{-\rho_{\zeta}(\mathbf{r})/\rho_{0}}].
\end{equation}
The parameter $\rho_{0}$ marks the density value above which non-ideal effects come into play. The prefactor $\rho_{0}$ in (\ref{PSI}) is used to ensure that for small densities the pseudopotential is linear in the density $\rho_{\zeta}$. Despite their inherent microscopic simplicity, the above dynamic rules are able to promote a host of non-trivial collective effects \citep{CHEM09,EPL10}.  The model gives direct access to the hydrodynamical variables, i.e. density and velocity fields, as well as the local (in time and space) stress tensor in the system, the latter characterised by both the viscous as well as the elastic contributions. When the strength parameter ${\cal G}_{AB}/\rho_0^2$ in the phase-separating interactions (\ref{Phase}) is chosen above a critical value, the model achieves phase separation and promotes the emergence of diffuse interfaces. The use of competing interactions (\ref{NNandNNN}) is instrumental to achieve a positive disjoining pressure $\Pi_d$ \citep{Colosqui13}. To quantify the emergence of the surface tension and the disjoining pressure, one has to consider a 1D problem. For a planar 1D interface, developing along $y$, the surface tension $\Gamma$ is a direct consequence of the pressure tensor developing at the non-ideal interface and is computed as the integral of the mismatch between the normal (N) and tangential (T) components of the pressure tensor. Such surface tension scales as as \citep{CHEM09}
\be\label{Gamma}
\Gamma =\int_{-\infty}^{+\infty} [P_{N}-P_T(y)] \, \D y \propto -\sum_{\zeta=A,B} \tilde{\cal G}_{\zeta \zeta} \int \left(\frac{\D \psi_{\zeta}}{\D y} \right)^2 \D y-\frac{{\cal G}_{AB}}{\rho_0^2} \int \frac{\D \rho_A}{\D y} \frac{\D \rho_B}{\D y} \, \D y.
\ee
The quantity $\tilde{\cal G}_{\zeta \zeta}={\cal G}_{\zeta \zeta ,1}+\frac{12}{7} {\cal G}_{\zeta \zeta ,2}$ comes from a proper combination of the coefficients in the competing interactions. For repulsive interactions, (${\cal G}_{AB}>0$) the second integral at the rhs is positive-definite, since $(\D \rho_A/\D y)(\D \rho_B/\D y) < 0$. With a proper use of the competing interactions, one can choose $\tilde{\cal G}_{\zeta \zeta}>0$, and the first term in the rhs of Eq.~\eqref{Gamma} is negative-definite; consequently, one can decrease the surface tension by simply increasing $\rho_0$. The decrease of the surface tension goes together with an increase of the disjoining pressure at the thin film interface. The emergence of a positive disjoining pressure $\Pi_d(h)$ can be controlled in numerical simulations by considering a thin {\it film} with two non-ideal flat interfaces, separated by the distance $h$. Following \citet{Bergeron}, we write the relation for the corresponding tensions
\be\label{FILMTENSION}
\Gamma_f(h)=2 \Gamma+\int^{\Pi_d(h)}_{\Pi_d(h=\infty)} h \, \D \Pi_d
\ee
where $\Gamma_f$ is the overall film tension. Similarly to what we have done for the surface tension $\Gamma$, the expression for $\Gamma_f$ is known in terms of the mismatch between the normal and tangential components of the pressure tensor \citep{Toshev,Derjaguin}, $\Gamma_f=\int_{-\infty}^{+\infty} [P_N-P_T(y)] \, \D y$,  where, in our model, $P_N-P_T(y)=p_s(y)$ can be computed analytically \citep{Shan08,SbragagliaBelardinelli13}. All the detailed expressions for the interaction stress tensor are reported in Appendix \ref{AppendixA}. From the relation $s(h)=\Gamma_f(h)-2 \Gamma$ it is possible to compute the disjoining pressure: a simple differentiation of $s(h)$ permits  to determine the first derivative of the disjoining pressure, $\D s(h)/\D h=h \,\D \Pi_d/\D h$.  This information, supplemented with the boundary condition $\Pi_d(h \rightarrow \infty) = 0$, allows to completely determine the disjoining pressure of the film \citep{Sbragaglia12}. In Fig.~\ref{fig:disjoining} we analyse quantitatively some of these features. In particular we consider the interaction parameters ${\cal G}_{AB}=0.405$, ${\cal G}_{\zeta \zeta ,1}=-9.0$, ${\cal G}_{\zeta \zeta ,2}=8.1$ with $\rho_0$ chosen in the interval $[0.72:0.84]$. All numbers are reported in lbu (lattice Boltzmann units). As we can see, by increasing the value of $\rho_0$, we enhance the energy barrier at the onset of the film rupture.

The body force $\mathbf{F}_b=\mathbf{F}_{P} + \mathbf{F}_{D}$ in Eq.~\eqref{LB} contains the driving due to the imposed (constant) pressure gradient ($\mathbf{F}_P$) and a drag force ($\mathbf{F}_D$) mimicking the friction between bubbles and confining plastic plates, as in the experimental setup (Fig.~\ref{Fig:setup_Benji}). Such drag force is taken to be proportional to the velocity vector, as in \citet{Janiaud06}, i.e. $\mathbf{F}_D = - \beta \mathbf{v}$. Once the droplets are stabilised with a positive disjoining pressure, different packing fractions and polydispersity of the dispersed phase can be achieved. In the numerical simulations presented in the following sections, the fraction of the continuous phase (i.e. the equivalent of the liquid fraction in the foam experiment) is kept approximately equal to $\phi_l \approx 7.5 \%$.
As already stressed in the introduction, the numerical model provides two basic advantages whose combination is not common. On one hand, it provides a realistic structure of the emulsion droplets, like for instance the Surface Evolver method \citep{Surface1,Surface2,Surface3}; at the same time, due to its built-in properties, the model gives direct access to dissipative mechanisms in thin films. This latter point will be further discussed and detailed in Appendix \ref{AppendixB}.  The viscous ratio between the dispersed phase and the continuous phase is kept fixed to $\chi=1$ (the simulation parameters are summarised in Tab.~\ref{Tab:summary_sim}). This choice is dictated by purely numerical reasons, as numerical instabilities emerge when one considers the case of a viscous ratio much smaller or much larger than unity. Nevertheless, we can use this as an advantage in our joint numerical and experimental study, as it offers the possibility to test the robustness of the experimental findings versus a change in the viscous ratio $\chi$ between the dispersed phase and the continuous phase. It is also comforting that the latest version of our GPU code \citep{GPU} allows for the simulation of emulsion droplets and their statistics in a reasonable amount of time. The current version runs on multiple-GPU and, by using a combination of CUDA {\em streams} and non-blocking MPI primitives, it is able to overlap completely the computation within the bulk of the domain with the exchange of the boundaries. Most simulations have been carried out on Kepler ``Titan'' GPUs, featuring 14 Streaming Multiprocessors, with a total of 2688 cores running at 0.88 Ghz and a memory bandwidth exceeding 200 GBytes/sec. Each run, spanning multi-million time steps for every single set of parameters, takes less than $12$ hours, to be compared with a running time of about $30$ hours on previous generation ({\em Fermi}) GPU cards. The speedup with respect to a highly tuned (multi-core) CPU version is above one order of magnitude. To develop a systematic analysis of plastic events, we perform a Voronoi tessellation (using the {\it voro++} libraries \citep{Rycroft06}) constructed from the centres of mass of the droplets, a representation which is particularly well suited to capture and visualise plastic events in the form of droplets rearrangements and topological changes, occurring within the material.


\begin{table}
  \centering
  \begin{tabular}{@{\extracolsep{\fill}}|c|c|c|c|c|c|}
  \hline
  RUN & $F_P$ & $|u_{wall}|$ & $v_0$ & $\beta$ & $\beta^*$ \\
      & lbu & lbu & lbu & lbu &  \\
  \hline
  P1 & $5 \times 10^{-7}$ & $0.0$ & $2.15 \times 10^{-2}$ & $0$ & $0$ \\
  P2 & $5 \times 10^{-7}$ & $0.0$ & $1.05 \times 10^{-2}$ & $10^{-5}$ & $100$\\
  P3 & $5 \times 10^{-7}$ & $0.0$ & $8.05 \times 10^{-3}$ & $2 \times 10^{-5}$ & $200$ \\
  P4 & $4 \times 10^{-7}$ & $0.0$ & $6.10 \times 10^{-3}$ & $2 \times 10^{-5}$ & $200$\\
  P5 & $3 \times 10^{-7}$ & $0.0$ & $4.55 \times 10^{-3}$ & $0$ & $0$\\
  P6 & $3 \times 10^{-7}$ & $0.0$ & $1.65 \times 10^{-3}$ & $2 \times 10^{-5}$ & $200$\\
  C1 & $0.0$ & $2 \times 10^{-2}$& $0.0$ & $0$ & $0$\\
  C2 & $0.0$ & $2 \times 10^{-2}$ & $0.0$ & $10^{-5}$ & $100$\\
  \hline
\end{tabular}
\caption{Summary of the simulations parameters. The first six rows refer to runs in the Poiseuille (P$\#$) flow setup, while the last two are relative to the Couette (C$\#$) flow numerical simulations. Other relevant parameters (kept fixed among the various runs) are the fraction of the continuous phase $\phi_l \approx 7.5 \%$ and the viscous ratio between the dispersed and continuous phase $\chi = 1$. The interaction parameters for the phase separating interactions (see Eq.~\eqref{Phase}) and competing interactions (see Eqs.~\eqref{NNandNNN}) are given in the text and the pseudo-potential reference density is $\rho_0 = 0.83$. The disjoining pressure for these interaction parameters is characterised in Fig.~\ref{fig:disjoining}. The total integration time is $T_{tot} = 2 \times 10^6$ lbu (lattice Boltzmann units) of which $T_{ss} = 1.25 \times 10^6$ steps in the steady state.}
  \label{Tab:summary_sim}
\end{table}


\section{Results and discussions}\label{sec:ResultsandDiscussion}

\subsection{Experimental velocity profiles} \label{Subsec:experimental_velocity_profiles}

\begin{figure}
\begin{center}
\includegraphics[scale=0.5]{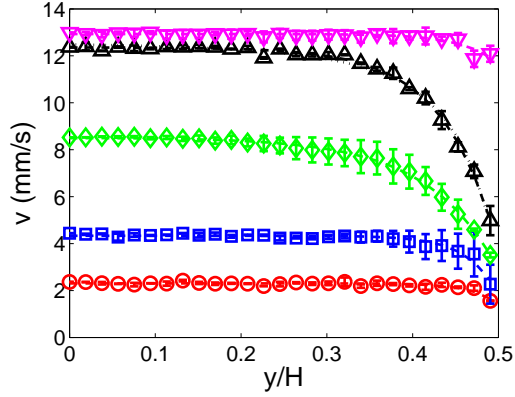}
\caption{Velocity profiles for the five experiments described in Tab.~\ref{Tab:summary_exp}. These data have been symmetrised with respect to the centreline; the error bars denote the standard deviation resulting from this averaging. Each dashed line is a fit of the data by the law (\ref{Eq:fit_velocity}); see Tab.~\ref{Tab:summary_exp} for the values of the best fitting parameters. The dotted line is a fit of the data series $\vartriangle$ with the law (\ref{Eq:fit_nonlinear_laws}) accounting for nonlinear friction and viscous stress; see Sec.~\ref{Subsubsec:nonlinear_model} and Appendix~\ref{AppendixC} for details. It is barely distinguishable from the dashed line.}
\label{Fig:Velocity_exp}
\end{center}
\end{figure}

The velocity profiles measured for five experiments are shown in Fig.~\ref{Fig:Velocity_exp}. They are quite flat at the centre ($y=0$) of the channel, although not completely flat as would be expected from a Herschel--Bulkley model, and decrease significantly close to the side walls ($y=\pm H/2$). They are well fitted by an exponential profile:
\begin{equation}\label{Eq:fit_velocity}
v(y) = v_1 \left( 1 + A\cosh\frac{y}{L_v} \right) = v_0 \frac{\cosh (H/2L_v) - \alpha_s - (1 - \alpha_s)\cosh (y/L_v)}{\cosh (H/2L_v) - 1} ,
\end{equation}
with a set of three fitting parameters: either $v_1$, $A$ and $L_v$, or $v_0$, $\alpha_s$ and $L_v$. We will retain the latter set of parameters, which has a clear physical meaning: $v_0 = v(y=0)$ is the centreline velocity, $\alpha_s = v(y = \pm H/2)/v_0$ is the relative slip, i.e. the ratio of the slip velocity to the centreline velocity. The parameter $L_v$, that we will henceforth call the {\it velocity localisation length}, describes the range of influence of the walls friction on the velocity profile. The values of the best fitting parameters are reported in Tab.~\ref{Tab:summary_exp}. Among the four experiments run at constant control parameters except the driving flow rate, the relative slip tends to decrease, and the localisation length to increase, at increasing flow rate, except the experiment at flow rate 102.5~mL/min. The fifth experiment is run at larger liquid fraction that the four other: it shows a larger relative slip, and a smaller localisation length, than the experiment with comparable flow rate.

\subsection{Comparison of experiments with a local model} \label{Subsec:local_model}

\subsubsection{Linear model} \label{Subsubsec:linear_model}

\begin{figure}
\begin{center}
\includegraphics[scale=0.5]{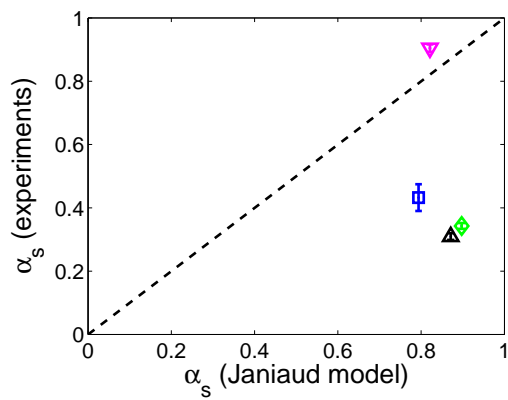}
\caption{Relative slip measured in the experiments described in Tab.~\ref{Tab:summary_exp}, as a function of the relative slip (\ref{Eq:slip_velocity_Janiaud_model}) predicted by the model of \citet{Janiaud06}.}
\label{Fig:Slip_velocity_exp_Janiaud}
\end{center}
\end{figure}

To provide analytical reference equations for the velocity profiles and place our work in the context of the existing literature, we start by comparing our velocity profiles to local models. We start by a comparison to the model of \citet{Janiaud06}. It is appealing, owing to its simplicity, and it has been shown to reproduce well experimental velocity profiles for foam flows in plane Couette geometry \citep{Katgert08}. The model considers a steady unidimensional flow, where inertia vanishes identically. We also neglect end effects, hence assume that flow is streamwise invariant. Hence, the flow profile writes: $\mathbf{v} = v(y)\mathbf{e}_x$, with $x$ the streamwise direction and $y$ the spanwise one. The streamwise invariance implies a constant pressure drop: $\boldsymbol\nabla P = \mathbf{e}_x \D P/\D x$ with $\D P/\D x$ constant. From momentum conservation, $ \mathbf{0} = \boldsymbol\nabla \cdot \boldsymbol\sigma - \boldsymbol\nabla P + 2\mathbf{f}_v/h$, with $\mathbf{f}_v$ the foam/wall friction force per unit area. Taking the streamwise component of the equation, we get:
\begin{equation}\label{Eq:general_equation_for_stress}
0 = \frac{\D\sigma}{\D y} - \frac{\D P}{\D x} + \frac{2}{h} f_v ,
\end{equation}
where $\sigma$ is the $xy$ component of the stress tensor. The model assumes that for the shear stress: $\sigma = \sigma_Y f_e(\gamma/\gamma_Y) + \eta\dot{\gamma}$ with $\gamma$ the shear strain and $\dot{\gamma} = \D v/\D y$ the shear rate, $\sigma_Y$ and $\gamma_Y$ the yield stress and the yield strain, respectively, $\eta$ a plastic viscosity of the foam, and $f_e$ a function quantifying the variation of the elastic stress with the shear strain. Inserting this model in (\ref{Eq:general_equation_for_stress}) yields:
$$
0 = \eta\frac{\D^2 v}{\D y^2} + \sigma_Y \frac{\D\gamma}{\D y} f_e' \left( \frac{\gamma}{\gamma_Y} \right) - \frac{\D P}{\D x} - \beta v ,
$$
where $f_v$ is assumed to be proportional to the velocity, and $\beta$ defined as:
\begin{equation}\label{Eq:fv_linear}
f_v = -\frac{1}{2} h\beta v .
\end{equation}
If we neglect the elastic term for simplicity, we get the following ODE for the velocity:
$$
\frac{\D^2 v}{\D y^2} - \frac{v}{L_0^2} = -\frac{v_1}{L_0^2} ,
$$
where:
\begin{equation}\label{Eq:definition_L0}
L_0 = \sqrt{\frac{\eta}{\beta}} .
\end{equation}
A first boundary condition comes from the fact that $x$ is a symmetry axis, hence $v$ is an even function of $y$, and we recover the exponential profile (\ref{Eq:fit_velocity}): $v(y) = v_1 [1 + A\cosh(y/L_0)]$, first proposed in Sec.~\ref{Subsec:experimental_velocity_profiles} as an empirical fit, with the characteristic velocity $v_1$ proportional to the pressure gradient:
\begin{equation}\label{Eq:definition_v0}
v_1 = -\dfrac{L_0^2}{\eta} \dfrac{\D P}{\D x} .
\end{equation}
As shown in Sec.~\ref{Subsec:experimental_velocity_profiles}, it turns out that this functional form, with $v_1$, $A$ and $L_0$ as free fitting parameters, reproduces very well the experimental profiles (Fig.~\ref{Fig:Velocity_exp}). However, there is a second boundary condition, coming from a force balance of the foam at the wall:
\begin{equation}\label{Eq:boundary_condition_side_wall}
\sigma = \pm f_v ~ \mathrm{at} ~ y = \pm H/2.
\end{equation}
A macroscopic, visible signature of the balance (\ref{Eq:boundary_condition_side_wall}) is the angle between the bubble edges and the side walls in Fig.~\ref{Fig:setup_Benji}b, see also \citet{DolletCantat10}. Inserting (\ref{Eq:fv_linear}), (\ref{Eq:definition_L0}) and (\ref{Eq:definition_v0}) in (\ref{Eq:general_equation_for_stress}), $\D\sigma/\D y = \beta(v - v_1)$, hence:
$$
\sigma\left( y = H/2 \right) = \beta v_1 A\int_0^{H/2} \cosh \frac{y}{L_0} ~\D y = \beta v_1 AL_0 \sinh \frac{H}{2L_0} ,
$$
which we insert in (\ref{Eq:boundary_condition_side_wall}) to get:
\begin{equation}\label{Eq:velocity_profile_Janiaud_model}
v(y) = v_1 \left[ 1 - \frac{h\cosh(y/L_0)}{2L_0 \sinh(H/2L_0) + h\cosh(H/2L_0)} \right] .
\end{equation}
This new functional form, with $v_0$ and $L_0$ as free fitting parameters, does not fit the experiments. Actually, there is a major discrepancy with the relative slip; setting $y = H/2$ in (\ref{Eq:velocity_profile_Janiaud_model}), we get:
\begin{equation}\label{Eq:slip_velocity_Janiaud_model}
\alpha_s = \frac{1}{1 + \frac{h}{2L_0} \coth \frac{H}{2L_0}} \simeq \frac{1}{1 + h/2L_0} ,
\end{equation}
since $\coth (H/2L_0)$ is very close to 1 for all our experiments. This prediction is much higher than the experimental value, except for the wet foam (Fig.~\ref{Fig:Slip_velocity_exp_Janiaud}).

\subsubsection{Nonlinear model} \label{Subsubsec:nonlinear_model}

A possible reason for the discrepancy lies in the fact that the friction force is nonlinear in velocity, and the viscous stress is nonlinear in shear rate. Following \citet{Denkov09}, the internal viscous stress for a 3D foam equals $\sigma = (\tau_{VF} + \tau_{VS})\Gamma/R$ (film and interface contribution respectively) with $R$ the bubble radius, and:
\begin{equation}\label{Eq:film_friction}
\tau_{VF} = 1.16\mathrm{Ca}_{\dot{\gamma}}^{0.47} (1 - \phi_l)^{5/6} \frac{(0.26 - \phi_l)^{0.1}}{\phi_l^{0.5}} ,
\end{equation}
with $\mathrm{Ca_{\dot{\gamma}}} = \mu\dot{\gamma}R/\Gamma$ the capillary number ($\mu = 10^{-3}$~Pa~s: bulk viscosity), and $\tau_{VS} = 9.8\pi B\dot{\gamma}^{0.18}$ with $B = 2.12\times 10^{-3}$~S.I. an empirical constant for SLES/CAPB/MAc foams. Using as orders of magnitude from the experiments $R\approx\sqrt{A/\pi} \approx 2$~mm and $\dot{\gamma} \approx v_0/L_v \approx 1$~s$^{-1}$, we get $\tau_{VS}/\tau_{VF} \approx 0.6$, hence the film term is dominant, although the surface term is not negligible. Keeping only the film term for simplicity, Eq.~(\ref{Eq:film_friction}) shows that the viscous stress scales sublinearly with the shear rate:
\begin{equation}\label{Eq:nonlinear_viscous_stress}
\sigma = \eta'\dot{\gamma}^{0.47} ,
\end{equation}
where the prefactor $\eta'$ (primed to distinguish it from the plastic viscosity in the linear law used in Sec.~\ref{Subsubsec:linear_model}) is:
\begin{equation}\label{Eq:plastic_viscosity}
\eta' = 1.16\frac{\mu^{0.47} \Gamma^{0.53}}{R^{0.53}} (1 - \phi_l)^{5/6} \frac{(0.26 - \phi_l)^{0.1}}{\phi_l^{0.5}} .
\end{equation}

For solutions giving rigid interfaces, like SLES/CAPB/MAc \citep{Golemanov08}, foam/wall friction is quantified by the force per unit area (or equivalently the wall stress) \citep{Denkov09}:
\begin{equation}\label{Eq:wall_stress}
f_v = \frac{\Gamma}{R} \left[ 1.25C_{IF} \sqrt{\mathrm{Ca}^*} \sqrt{\frac{F_3}{1-F_3}} + 2.1C_{IL} (\mathrm{Ca}^*)^{2/3} \right] F_3 ,
\end{equation}
with two empirical constants $C_{IF} = 3.7$ and $C_{IL} = 3.5$, $\mathrm{Ca}^* = \mu v/\Gamma$ another capillary number, and:
$$ F_3 = \sqrt{1 - 3.2\left( \frac{1 - \phi_l}{\phi_l} + 7.7 \right)^{-1/2}} . $$
For $v \approx 1$~cm/s, the ratio of the second term to the first term in (\ref{Eq:wall_stress}) is 5, hence we neglect the second term. Eq.~(\ref{Eq:wall_stress}) then shows that the friction force scales sublinearly with the velocity:
\begin{equation}\label{Eq:nonlinear_friction}
f_v = -\frac{1}{2} h\beta'\sqrt{v} ,
\end{equation}
with the following value of the friction constant (primed to distinguish it from its counterpart in the linear law used in Sec.~\ref{Subsubsec:linear_model}):
\begin{equation}\label{Eq:parameter_beta}
\beta' = \frac{2.5C_{IF}}{hR} \sqrt{\Gamma\mu} \sqrt{\frac{F_3^3}{1-F_3}} .
\end{equation}
Like in Sec.~\ref{Subsubsec:linear_model}, see Eq.~(\ref{Eq:definition_L0}), we can construct a characteristic length from $\eta'$ and $\beta'$. To do so, it is convenient to replace the exponent 0.47 by $1/2$ in (\ref{Eq:nonlinear_viscous_stress}), recasting the factor $\dot{\gamma}^{0.03}$ in the definition (\ref{Eq:plastic_viscosity}) of $\eta'$; this factor is almost constant, and equal to 1, for all our experiments. The characteristic length is then $Y = (\eta'/\beta')^{2/3}$. We compute with all the experimental values of the parameters appearing in (\ref{Eq:plastic_viscosity}) and (\ref{Eq:parameter_beta}): $Y = 1.9$~mm for $\phi_l = 4.8\%$, and 2.7~mm for $\phi_l = 16.9\%$. These orders of magnitude are compatible with the experimental values of the localisation lengths (Tab.~\ref{Tab:summary_exp}).

The effect of nonlinear friction and viscous stress on the velocity profile has been theoretically considered for Couette flows \citep{Weaire08,Weaire09}, but not for Poiseuille flows. Therefore, in Appendix~\ref{AppendixC}, we compute analytically the velocity profile using the nonlinear laws (\ref{Eq:nonlinear_viscous_stress}) and (\ref{Eq:nonlinear_friction}), and we show that the role of these nonlinearities on the velocity is negligible.

\begin{figure}
\begin{center}
\includegraphics[scale=1]{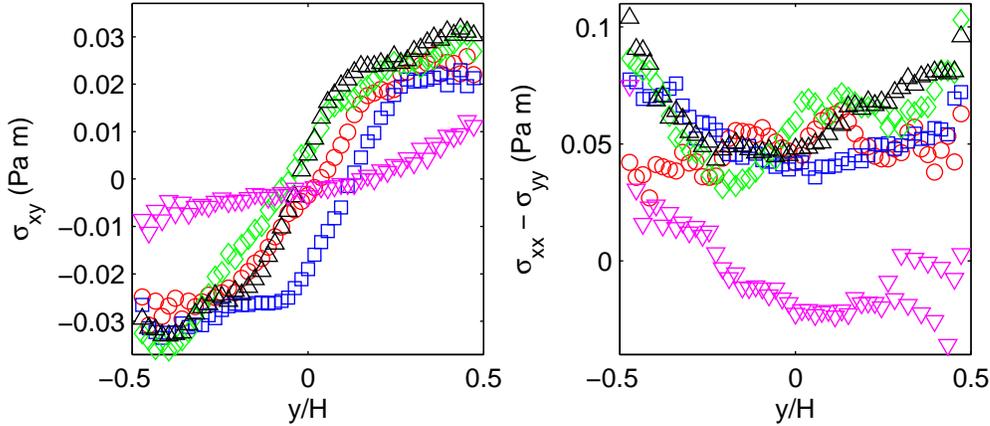}
\caption{Left panel: shear component of the elastic stress, and right panel: normal difference of the elastic stress, for the five experiments described in Tab.~\ref{Tab:summary_exp}.}
\label{Fig:Shear_elastic_stress_exp}
\end{center}
\end{figure}

Hence, the model of \citet{Janiaud06} is too simple to model wall slip in our experiments, which suggests that the role of elastic stresses is crucial. This is qualitatively supported by the fact that the only experiment for which the local model is quite accurate in predicting the amount of slip is for a wet foam, which stores less elastic energy \citep{Cantat13}. To further support this idea, we plot the shear component of the elastic stress and the normal elastic stress difference in Fig.~\ref{Fig:Shear_elastic_stress_exp}. The shear elastic stress is indeed about four times weaker for the wet foam than for the four other experiments. For these experiments at given bubble area and liquid fraction, its variation across the channel is as follows: towards the centre of the channel, although with a significant asymmetry for some experiments, there is a zone of quasilinear increase around $\sigma_{xy} = 0$. The width of this region decreases slightly at increasing flow rate. Outside this region, the elastic shear stress plateaus to a value which does not depend much on the flow rate. Interestingly, there is still some velocity variation, and a significant plastic activity, outside those regions where the shear elastic stress plateaus. Except the experiment for the wet foam, the normal elastic stress difference $\sigma_{xx} - \sigma_{yy}$ is always positive, i.e. the bubbles are elongated streamwise, an effect which is clearly visible on Fig.~\ref{Fig:setup_Benji}b. It tends to increase towards the wall.

Eq.~(\ref{Eq:general_equation_for_stress}) expresses the balance between the driving pressure gradient, the foam/wall friction, and the gradient of elastic and viscous stresses. Close to the middle of the channel, the velocity gradient is very weak, hence the viscous stress is negligible, and the gradient of the shear elastic stress is roughly constant (Fig.~\ref{Fig:Shear_elastic_stress_exp}). It is interesting to compare the value of this gradient $\D\sigma_{xy}/\D y$ and the pressure gradient $\D P/\D x$. Their experimental values are reported in Tab.~\ref{Tab:summary_exp}; the pressure gradient is always larger than the gradient of shear elastic stress, the missing part being the friction. This is a major difference with Poiseuille experiments in 3D channels \citep{Goyon08,Goyon10,Geraud13}, where the friction force is absent. This prevents us from measuring directly the spanwise stress from the pressure gradient, contrary to the aforementioned studies.

\subsection{Numerical simulations and comparison with the fluidity model}\label{sec:numexp}

Having shown the inaccuracies of a local model without elasticity to capture our experimental data thanks to the inspection of the boundary condition at the wall, we now want to test the effect of elasticity. Local visco-elastoplastic models could be used \citep{Cheddadi12}, but it is not straightforward to deduce from them testable predictions. We thus test a kinetic elastoplastic model \citep{Bocquet09}, which encompasses the effect of elasticity through the nonlocal relaxation of elastic stress induced by plastic events. It predicts that the rate of plastic events should be proportional to the fluidity $\dot{\gamma}/\sigma$, a prediction that we can readily test. However, there is a difficulty in testing this nonlocal model against our experiments. The role of friction is crucial in experiments, whereas the nonlocal model has been set up and tested in its absence, although recent studies have considered the coupled role of friction and nonlocality \citep{Barry11}. Indeed, friction complicates the stress profile across the channel, as discussed in Sec.~\ref{Subsec:local_model}, and it is thus not straightforward to extract relevant flow curves $\sigma(\dot{\gamma})$ from our experiments. Hence, it is interesting to run numerical simulations, where the friction can be set off and tuned at will.

\begin{figure}
\begin{center}
\includegraphics[scale=0.7]{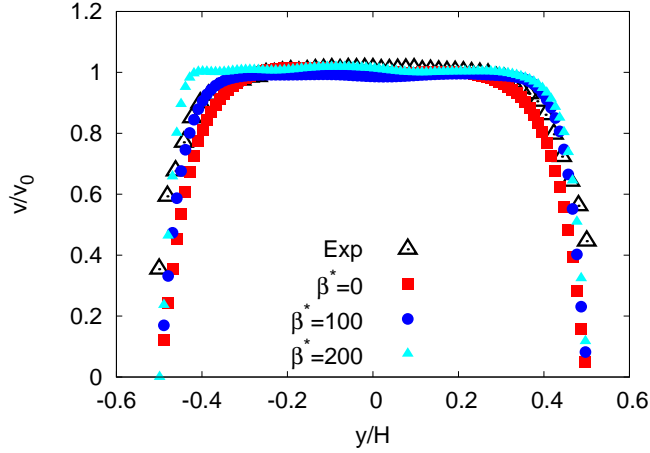}
\caption{Numerical velocity profiles, normalised by the centreline velocity $v_0$, at changing the friction constant. Data from three sets of simulations are shown with dimensionless friction parameter (see equation \ref{betastar}) $\beta^*=0,100,200$ (runs P1-3 in table Tab.~\ref{Tab:summary_sim}); the experimental velocity profile with flow rate 152.5 ml/min (see Tab.~\ref{Tab:summary_exp}) is also reported  to show that we are able to tune the friction parameters in the numerics to achieve the same localisation observed in the experiments. The equivalent $\beta^*$ in the experiments is $\beta^* \approx 250$ (see text for details). On the abscissae, the $y$-location across the channel has been normalised by the total channel height.}
\label{Fig:velocity_field_Stress}
\end{center}
\end{figure}

Various sets of numerical simulations have been performed in the $(F_P, \beta^*)$ parameter space (see Tab.~\ref{Tab:summary_sim} for the numerical values used), where $\beta^*$ is meant to be the value of $\beta$ made dimensionless with the channel width $H$ and viscosity $\eta$, i.e. \be\label{betastar}
\beta^{*}=\frac{\beta H^2}{\eta}=\frac{H^2}{L_0^2} ,
\ee
where the last equality is based on the definition of the localisation length $L_0$ given in (\ref{Eq:definition_L0}). A flat velocity profile in the bulk is shown by all curves (Fig.~\ref{Fig:velocity_field_Stress}), including the case with $\beta^*=0$, witnessing the presence of a non trivial bulk rheology \citep{Goyon08,Goyon10,Geraud13}. We also report the experimental velocity profile with flow-rate 152.5 ml/min (see Tab.~\ref{Tab:summary_exp}), just to show that we are able to tune the friction parameters in the numerics to achieve the same localisation observed in the experiments for which an equivalent friction parameter $\beta^* \approx 250$ can be estimated based on the friction constant (\ref{Eq:parameter_beta}) and the plastic viscosity (\ref{Eq:plastic_viscosity}). A direct comparison of the velocity profiles of experiments and numerical simulations is somehow irrelevant, because of the boundary conditions: as observed in the experimental data of Fig.~\ref{Fig:Velocity_exp}, slippage is found to occur at the surfaces of the experiments, while the numerical simulations are performed by imposing no-slip at the walls. Boundary conditions have an impact on the microscopic dynamics at the wall \citep{Mansard14}. The numerical simulations will be, on the other hand, quite useful in validating the picture of the plastic flow (Sec.~\ref{Subsubsec:plasticflow}) at changing the friction constant, a feature that is freely tunable in the numerics.

\begin{figure}
\begin{center}
\includegraphics[scale=0.6]{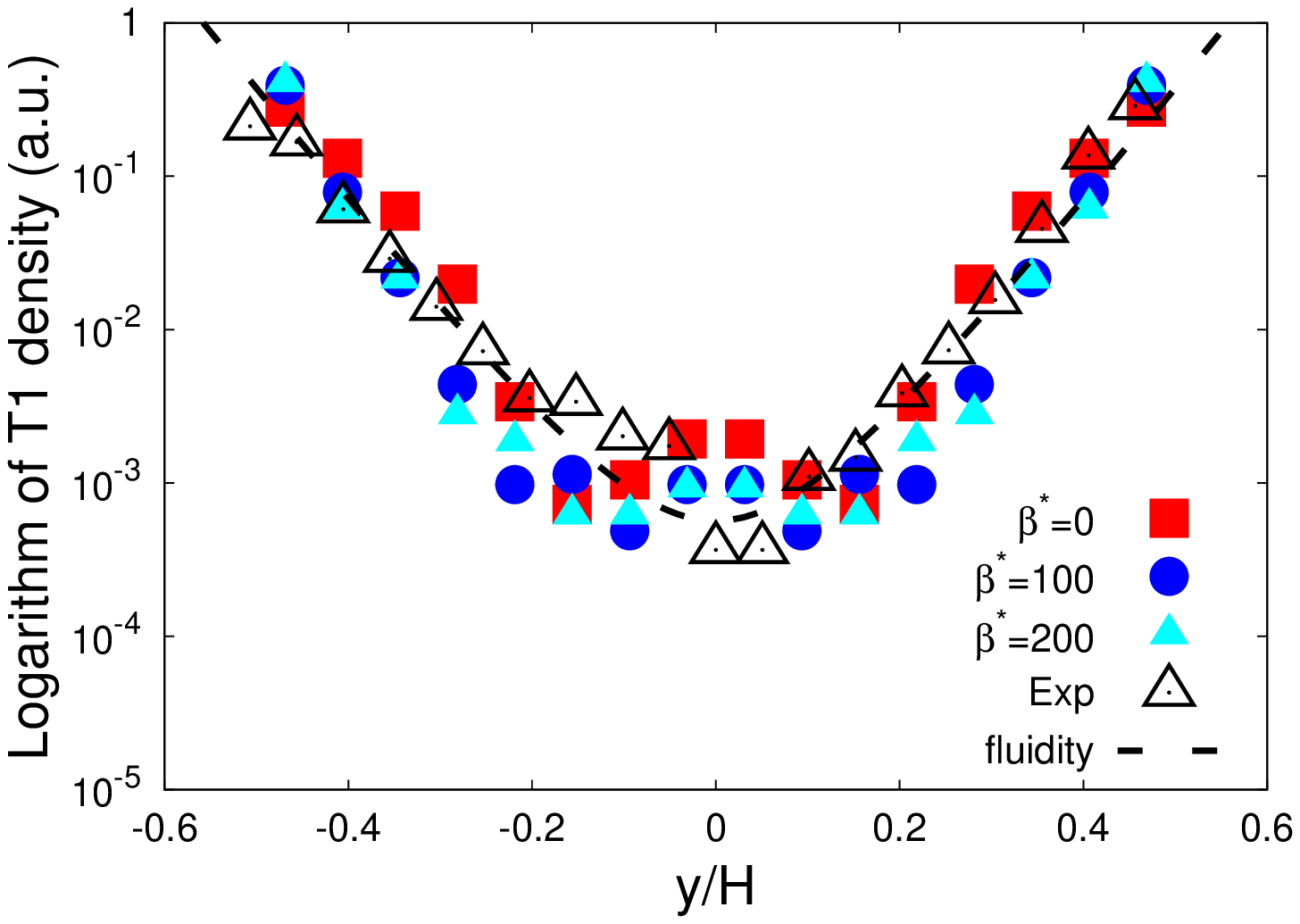}
\includegraphics[scale=0.6]{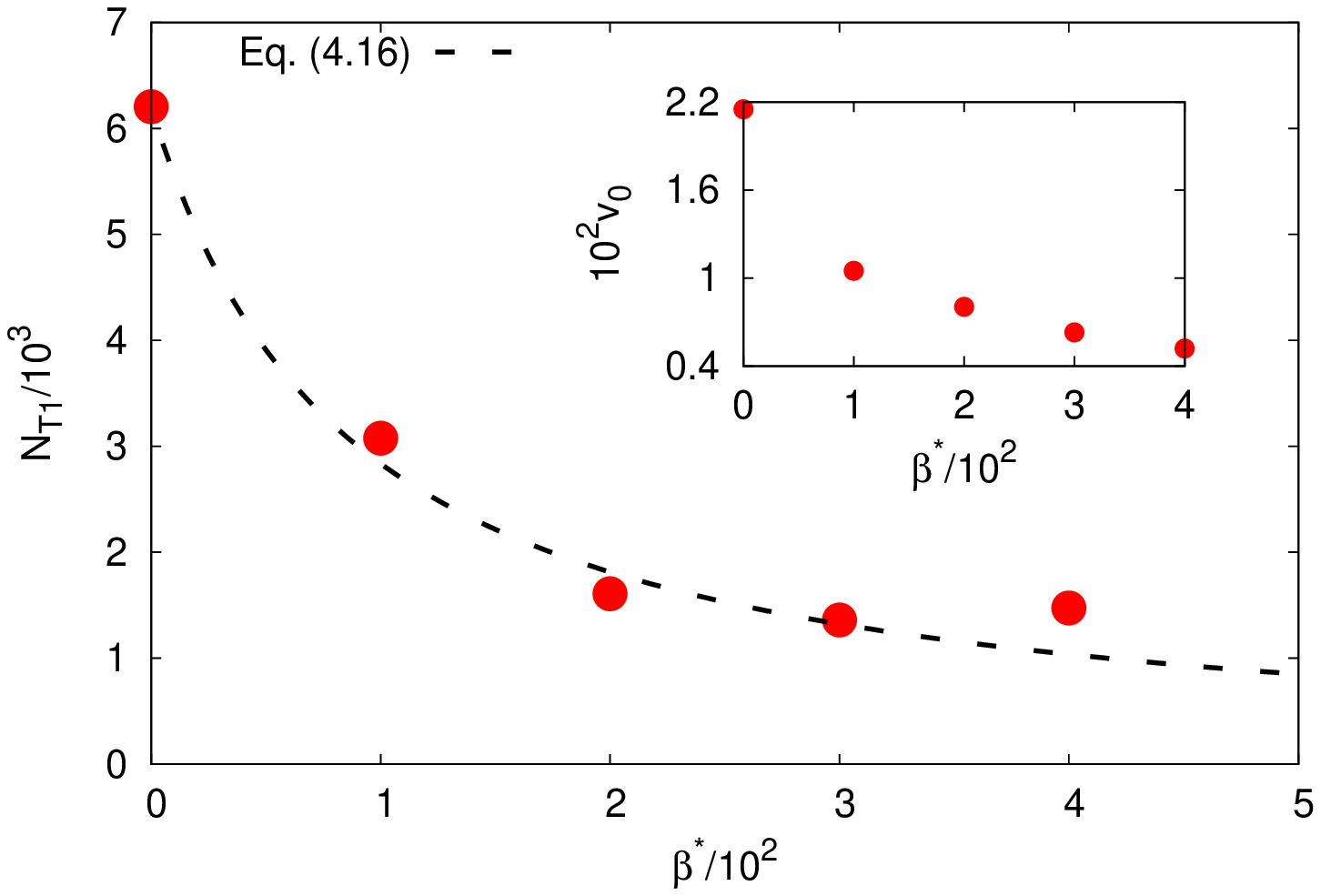}
\caption{Top Panel: plot of the rate of plastic rearrangements as a function of $y$: experiments ($\vartriangle$ in Tab.~\ref{Tab:summary_exp}) are compared with numerical data from runs P1-3. The dashed line indicates the function $\sinh(y/L_v)/y$, representing the fluidity profile based on the hyperbolic cosine fit of the velocity profile (see Sec.~\ref{Subsec:experimental_velocity_profiles} for details). Numerical data have been symmetrised. Bottom Panel: Total number of T1s as function of $\beta^*$ from the simulations; the dashed line is a fit with the functional form given in Eq.~\eqref{Eq:decay}. Inset: the centreline velocity $vs$ $\beta^*$ is reported.}
\label{Fig:T1_field_exp}
\end{center}
\end{figure}

The top panel of Fig.~\ref{Fig:T1_field_exp} indeed provides some indications that the extra friction (i.e. the confining plates) does not seem to dramatically affect the distribution of plastic events. There we plot the rate of plastic rearrangements, normalised by the total number of events, from experiments and numerics (for the three $\beta^*$'s). Data show a moderately good collapse onto each other. At a given driving pressure drop, increasing wall friction results in a decrease of the total number of plastic events $N_{T_1}$. We could estimate the number $N_{T_1}$ in the numerical simulations and it is reported in the bottom panel of Fig.~\ref{Fig:T1_field_exp}: for the same simulation time (see caption of Tab.~\ref{Tab:summary_sim}) plastic events diminish from $N_{T_1} \sim 6 \times 10^3$ to $N_{T_1} \sim 2 \times 10^3$  for increasing $\beta^*$ from $0$ to $400$. A similar trend is observed for the centreline velocity which is reported in the inset of the bottom panel of Fig.~\ref{Fig:T1_field_exp}. To make this statement more quantitative, we notice that the overall decrease in the number of plastic events can be well captured by the function
\be\label{Eq:decay}
G(\beta^*) = \frac{2 b^2 N_{T1}^{(0)}}{\beta^*}\left[1-\frac{1}{\cosh (\sqrt{\beta^*}/b)} \right] ,
\ee
a scaling behaviour that can be obtained from the expression of the centreline velocity, $v_0$ in (\ref{Eq:fit_velocity}), with $\alpha_s=0$ (no-slip boundary condition for the numerics). The parameter $N_{T1}^{(0)}$ in equation (\ref{Eq:decay}) sets the number of plastic events in the limit $\beta^* \rightarrow 0$ whereas the argument  $\sqrt{\beta^*}/b$ of the hyperbolic function in Eq.~\eqref{Eq:decay} is inversely proportional to the localisation length. The choice of equation (\ref{Eq:decay}) as a fitting function is suggested by the consideration that the total number of plastic events is dominated by events occurring in boundary regions where the shear stress is approximately constant and, hence, the fluidity $f$ is basically proportional to the shear rate $\dot{\gamma}$.
Consequently, being the number of events, by definition, equal to the integral of the corresponding rate $\Gamma_p$ and since
$\Gamma_p \propto f \approx |\dot{\gamma}|$, we can assume that $G(\beta^*) \propto \int_0^{H/2} |\dot{\gamma}| dy$, which equals the centreline velocity.
  Interestingly, the estimate of $b$ that we get from a best fit procedure ($b \approx 6.0$) is greater than the estimate of $b$ based on the friction localisation $L_0$ in (\ref{Eq:definition_L0}), which would yield $b=2$. This is an indication that the localisation lengh in the numerical simulations is larger than the localisation length induced by the extra friction $\mathbf{F}_D = - \beta \mathbf{v}$. This is not a surprise, because our numerical simulations have already confirmed the presence of a cooperativity length scale \citep{Sbragaglia12} without wall friction. This supports the idea that an effective localisation length results from the sum of the friction localisation plus cooperativity length \citep{Barry11}, an issue that we will further explore in Sec.~\ref{Subsubsec:plasticflow}.

The study of the spatial distribution in the number of plastic events and the simultaneous analysis of the localisation in the velocity profiles, allows to bridge between the ``microscopic'' details of local irreversible plastic rearrangements and the macroscopic flow. It has been suggested that such a bridge can be established by introducing a cooperativity scale which determines correlations (nonlocal effects) in the flow rheology \citep{Goyon08,Goyon10,Geraud13}. The underlying idea is that correlations among plastic events exhibit a complex spatio-temporal scenario: they are correlated at the microscopic level with a corresponding cooperativity flow behaviour at the macroscopic level. The cooperativity scale $\xi$ \citep{Goyon08}, directly affects the fluidity, which has been claimed to follow a nonlocal diffusion equation where the diffusivity is directly proportional to $\xi^2$ \citep{Goyon08,Bocquet09}
\be\label{eq:fluidity}
\xi^2 \Delta f(\mathbf{x})+ f_b(\sigma(\mathbf{x}))-f(\mathbf{x}) =0.
\ee
The quantity $f_{b}$ is the bulk fluidity, i.e. the value of the fluidity in the absence of spatial cooperativity ($\xi=0$). The nonlocal equation (\ref{eq:fluidity}) has been justified \citep{Bocquet09} based on a kinetic model for the elastoplastic dynamics of a jammed material, which takes the form of a nonlocal kinetic equation for the stress distribution function. In the steady state, under the hypothesis of low cooperativity, the model predicts rheological equations which take the form of equation (\ref{eq:fluidity}), plus an equation predicting a proportionality between the fluidity and the rate of plastic events, a prediction that is interesting to test. A connection between the rate of plastic events and the fluidity field is indeed visible in the top panel of Fig.~\ref{Fig:T1_field_exp}. The dashed line indicates $\sinh(y/L_v)/y$, which is the ``synthetic'' fluidity profile (up to an unessential numerical scaling factor) based on the hyperbolic cosine fit of the velocity profile (see Sec.~\ref{Subsec:experimental_velocity_profiles}) and a linear variation of the shear stress across the channel. Interestingly, a significant plastic activity remains towards the centre of the channel, and it is well correlated to the fluidity field, which remains finite in such regions, whereas the strain-rate goes to zero. Moreover, a closer inspection reveals that the decrease in the number of plastic events is affected by the friction constant $\beta^*$, with a steeper decrease associated with the larger $\beta^*$. Fig.~\ref{Fig:T1_field_exp} calls therefore for a deeper understanding with regard to the link between the rate of plastic events and the local flow properties.

\subsection{Localisation lengths: comparison of plasticity and shear rate}\label{Subsubsec:plasticflow}

\begin{figure}
\begin{center}
\includegraphics[scale=0.6]{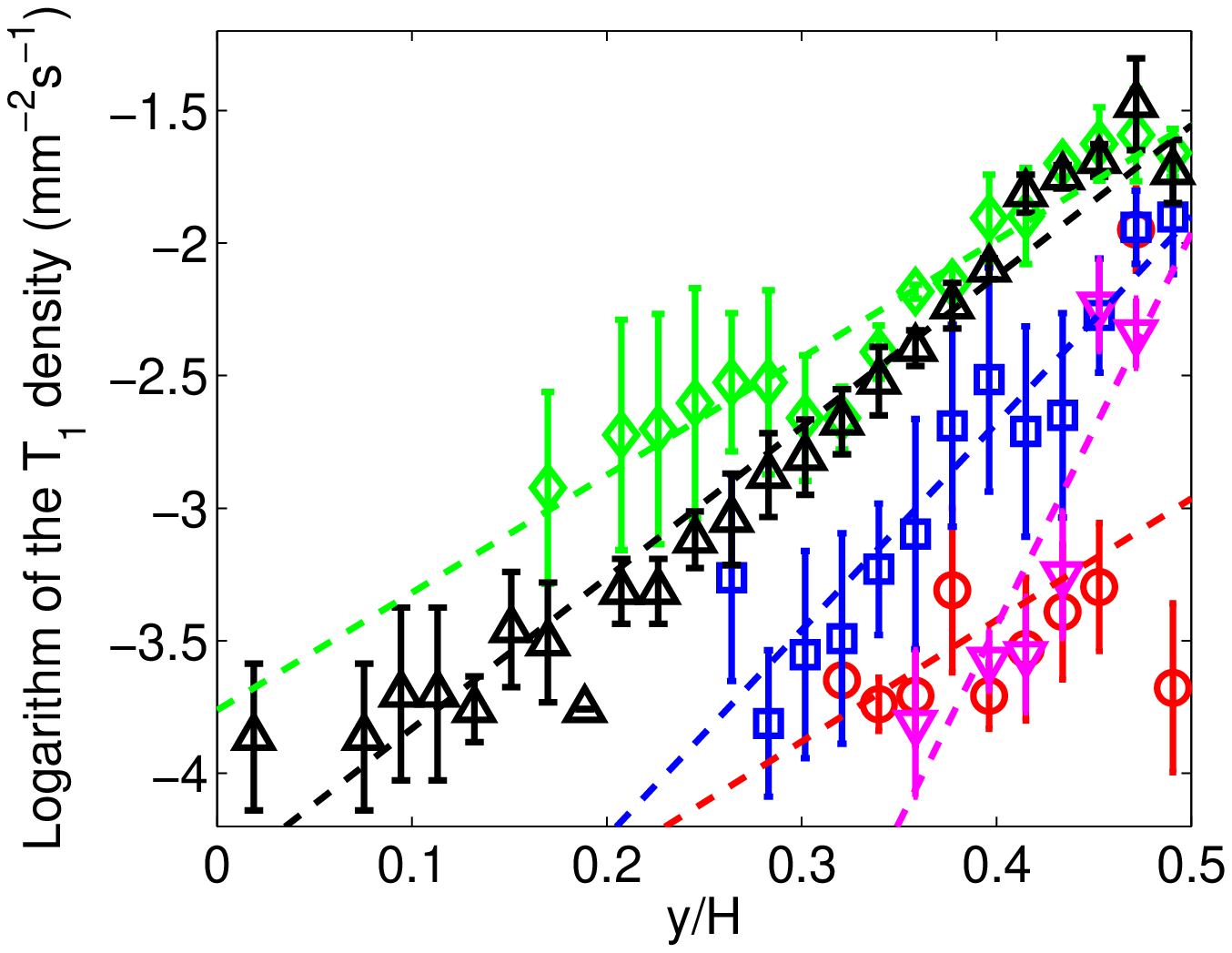}
\includegraphics[scale=0.6]{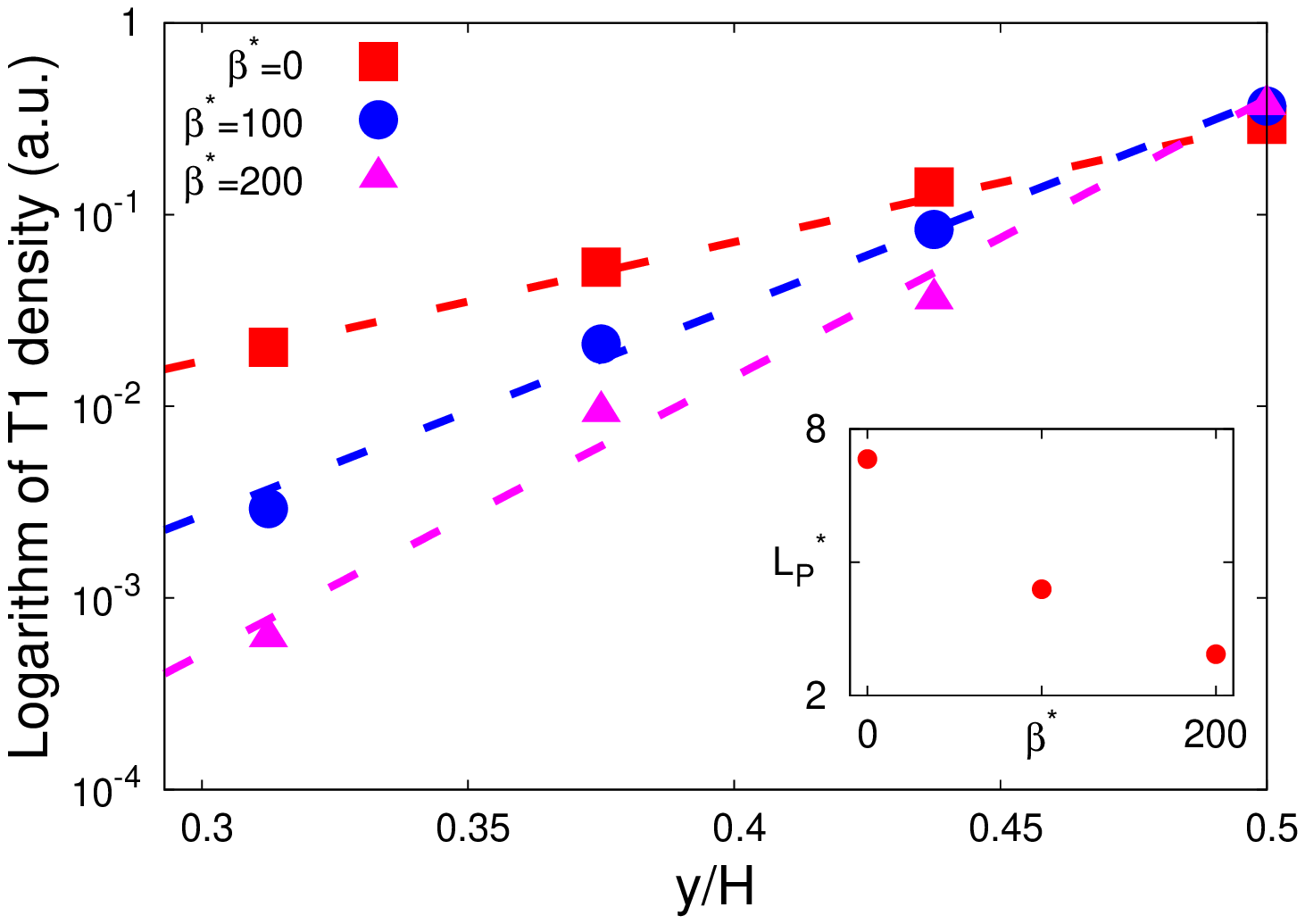}
\caption{Top Panel: Decimal logarithm of the density of T1s per unit time and area as a function of $y/H$, for the five experiments described in Tab.~\ref{Tab:summary_exp}. These data have been symmetrised with respect to the centreline; the error bars denote the standard deviation resulting from this averaging. Data below $10^{-4}$~mm$^{-2}\cdot$s$^{-1}$ are not shown because they are statistically irrelevant (less than ten T1s counted per bin per experiment). Dashed lines represent the linear fits of the data. Bottom Panel: data from numerical simulations complement the experimental results reported in the top Panel. In particular, to appreciate the effect of the friction at fixed pressure gradient, we show the Log-lin plot of the rate of plastic events from simulations P1-3 (fixed pressure drop and different $\beta^*$) close to the bottom wall (the dashed lines represent best linear fits of the data). Inset: Plastic localisation length as function of the friction parameter $\beta^*$.}
\label{Fig:Velocity_T1}
\end{center}
\end{figure}

To go further, we choose to explore the connection between the rate of plastic events and the local flow properties, by looking at the relationship between the localisation length of the velocity profiles, $L_v$, and the localisation length of the number of plastic events, $L_p$. This connection enables to compare experiments and simulations, despite their different boundary conditions. The localisation length of the velocity profiles $L_v$ is estimated by a hyperbolic cosine function $\cosh(y/L_v)$, from which the decay length $L_v$ is extracted (see Sec.~\ref{Subsec:experimental_velocity_profiles}). Simultaneously, the plastic localisation length $L_p$ is computed out of an exponential fit of the symmetrised rate of plastic events across the channel (Top Panel of Fig.~\ref{Fig:Velocity_T1}). Since our numerical simulations have already confirmed the presence of a cooperativity length scale \citep{Sbragaglia12} without wall friction, they are good candidates to complement the experimental findings, showing how the spatial distribution of plastic events is affected by a change in the friction $\beta$. Hence, in the Bottom Panel of Fig.~\ref{Fig:Velocity_T1} we also look at the localisation in the numerics, by fixing the pressure gradient and changing $\beta^*$, something that cannot be easily done in experiments with the data at hand. At fixed pressure gradient, we show the Log-lin plot of the rate of plastic events from simulations with different $\beta^*$. The extracted $L_p$ is found to be a decreasing function of $\beta^*$.

\begin{figure}
\begin{center}
\includegraphics[scale=0.6]{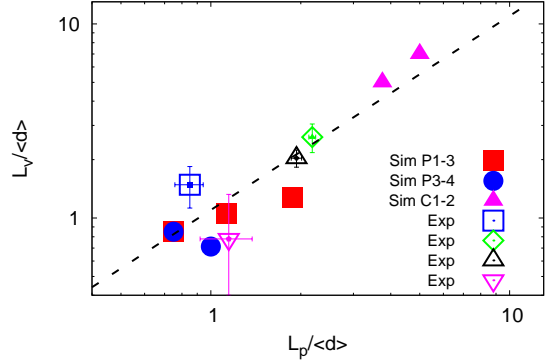}
\caption{Scatter plot of the shear localisation length $L_v$ (computed from a hyperbolic cosine fit of the velocity profiles) vs the plastic localisation length $L_p$ (computed out of an exponential fit of the symmetrised rate of plastic events across the channel) for three sets of data: experiments (symbols as in Tab.~\ref{Tab:summary_exp}), simulations of Poiseuille flow with fixed pressure drop and various $\beta^*$'s (filled squares) and with fixed $\beta^*=200$ and various pressure drops (filled circles) and simulations of Couette flow at two $\beta^*$'s (filled triangles); both lengths are normalised by the mean bubble diameter. The dashed line is the $L_v = L_p$ curve.}
\label{Fig:master-curve}
\end{center}
\end{figure}

In Fig.~\ref{Fig:master-curve} we report a scatter plot of the shear localisation length $L_v$ {\em versus} the plastic localisation length $L_p$ for three sets of data: experiments (symbols as in Tab.~\ref{Tab:summary_exp}), simulations with fixed pressure drop and various $\beta^*$'s (filled squares) and simulations with fixed $\beta^*=200$ and various pressure drops (filled circles). Fig.~\ref{Fig:master-curve} shows that the two localisation lengths are indeed close to each other. The fact that the values $L_p$ and $L_v$ agree, confirms the picture of the ``plastic flow"; it is also compatible with the fact that the rate of plastic events and the fluidity seem to be proportional (Sec.~\ref{sec:numexp}). More precisely, since the localisation length $L_v$ is much smaller than the channel width in our experiments and simulations, the stress does not vary much across the localisation zone, hence in most of the channel there is a good correlation between the shear rate and the rate of plastic events. \citet{Barry11} have combined the local model presented in Sec.~\ref{Subsec:local_model} with a nonlocal constitutive equation for the fluidity field, in the case of a Couette flow with linear laws for the viscous stress and the friction (as in Sec.~\ref{Subsubsec:linear_model}). They predicted that the localisation length of the velocity profile is an increasing function of both the cooperativity length $\xi$, and of the friction length $L_0$ defined by (\ref{Eq:definition_L0}). Here, this theoretical prediction can be tested for the first time \emph{versus} our experiments and simulations. Some care is required in doing so, because of the nonlinear nature of friction and viscous stress in experiments and of the difference between Couette and Poiseuille flows. However, we have shown in Appendix~\ref{AppendixC} that the effect of nonlinear friction and viscous stress is very weak. Since the localisation length $L_v$ is much smaller than the channel width in our experiments and simulations, the comparison with the Couette predictions is relevant. For this reason, we also repeated some numerical simulations in a Couette flow geometry (see Tab.~\ref{Tab:summary_sim}). The associated data nicely collapse on the same master curve, stressing even more the robustness of our findings at changing the load conditions. The simulations with a Poiseuille flow (inset of the Bottom panel of Fig.~\ref{Fig:T1_field_exp}) show that the localisation length $L_p$ is a decreasing function of $\beta$ at fixed pressure gradient. This is compatible with the local model, more precisely with the expression (\ref{Eq:definition_L0}) of the friction length. However, the latter tends to diverge at vanishing friction, whereas the localisation length remains finite in this limit; this suggests that the model of \citet{Barry11} breaks down at vanishing friction.

In experiments, for the foam of given liquid fraction and bubble area, the localisation length $L_v$ increases with increasing flow rate. The friction length $L_0$ being fixed, our result may suggest that the cooperativity length could increase with the flow rate.  However, looking at the signature of individual plastic events in elastic stress (or displacement) redistribution in their surroundings, we could not find a conclusive signature of an increasing range.  This question thus remains open and should be further addressed in the future. On the other hand, Tab.~\ref{Tab:summary_exp} shows that at given flow rate (up to 5\%), the localisation length is lower for a wet foam than for a dry one. This is in qualitative agreement with the experiments of \citet{Goyon10} for emulsions, who showed that the cooperativity length is an increasing function of the packing fraction above the jamming transition. We remark that testing the effect of the cooperativity length $\xi$ is highly nontrivial; whereas it has been shown to increase towards the jamming transition \citep{Bocquet09,Jop12,NicolasBarrat13}, the influence of shear rate remains rather unclear. In other experiments on concentrated emulsions \citep{Goyon08,Bocquet09}, a unique length $\xi$ was found to account for all experimental data for the flow profiles and local flow curves for a given volume fraction of the emulsion, independently of the pressure drop, confinement and surface nature. However, the flow behaviour in the limit of low shear rates is difficult to access experimentally. These conclusions apply equally well in the case of our numerical simulations where, for a given friction constant, we could not find any signature of an increase of the cooperativity length at changing the shear rate. However, also for the numerical simulations, the situations with low shear are very difficult to analyse, as they require a very large statistic. In some sense, the work presented here bypasses the problem of an accurate measurement of the cooperativity length close to the yield stress and directly explores the link between localisation phenomena in the velocity profiles and the micro-dynamics, characterised by the rate of plastic events and their localisation.\\

\subsection{Orientation of the plastic events}\label{ref:orientation}

\begin{figure}
\begin{center}
\includegraphics[width=15cm,keepaspectratio]{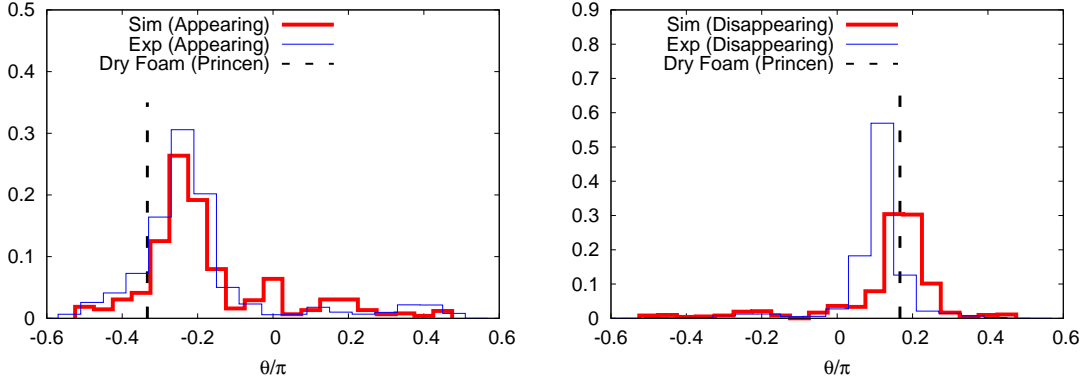}
\caption{Normalised distributions of the orientations of plastic events.  Distributions of appearing (left Panel) and disappearing (right Panel) centre-to-centre links of bubbles involved in rearrangements are shown, from experiment (thin blue line) and simulations (thick red line). $\theta$ denotes the angle formed by the link and the direction of the flow, i.e. the positive $x$-axis. Following \citet{Princen83}, the extreme values are found for a dry foam at liquid fraction $\phi_l = 0$ and are indicated with a dashed line.}
\label{Fig:T1_orientation}
\end{center}
\end{figure}

The importance of plastic rearrangements has been stressed in that the occurrence of these events induces long range correlations within the soft-glassy material. It is also acknowledged \citep{Picard04,Schall07} that T1s possess a non trivial angular structure with a quadrupolar topology. It seems, then, reasonable to argue that for a full understanding of the way they determine nonlocal effects inside the system, not only  the distribution of their locations in space, but also their orientational properties need to be addressed. Therefore, we go further with the description of plastic events, and study their angular statistics from experiments and simulations.   More precisely, focusing on the four bubbles involved in a T1, we define as a {\em disappearing} link the segment connecting the centres of the two bubbles which were in contact before the event (and which are then far apart), and as an {\em appearing} link the connector between the other two bubble centres (see also Sec.~\ref{sec:experiments} and Figs.~\ref{Fig:sketch_T1} and \ref{Fig:sketch_T1_sim}); we then measure for each event the angle between such links and the flow direction. We have observed that the angles are reversed between both sides of the channel, consistently with the fact that $y = 0$ is an axis of symmetry. Therefore, we choose to analyse the statistics of the quantities $\theta_d = \theta'_d \,\mathrm{sign}\,(y)$ and $\theta_a = \,\theta'_a \mathrm{sign}\,(y)$ (see Fig.~\ref{Fig:setup_Benji} for the sign convention of $y$). We did not observe a significant variation of the distribution of these angles across the channel, hence we analyse the distributions of these angles for all T1s, whatever their location across the channel. Fig.~\ref{Fig:T1_orientation} shows the histogram of $\theta_d$ and $\theta_a$ for one experiment and one simulation, while the average and standard deviation of these quantities are summarised for all experiments in Tab.~\ref{Tab:summary_exp}. This analysis shows that T1s have preferential orientations: $\theta_d$ is peaked around 0.5~rad, with a small dispersion, and $\theta_a$ around $-0.7$~rad, with a larger dispersion. The average values do not depend significantly on the flow rate. For the wet foam, $\theta_d$ is larger, and $\theta_a$ slightly smaller.


\begin{figure}
\begin{center}
\includegraphics[width=10cm,keepaspectratio]{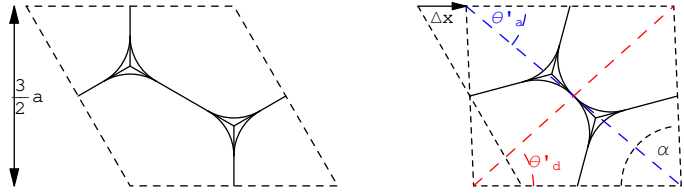}
\caption{Portion of an unsheared (left) and sheared (right) hexagonal foam. The strain is defined as $\gamma = 4\Delta x/3a$.}
\label{Fig:Princen_T1_orientation}
\end{center}
\end{figure}

We now derive some reference values for these angles from a microstructural analysis. Since our foams are rather monodisperse, it is interesting to use the simple geometrical model of a sheared 2D hexagonal foam \citep{Princen83} (see also \citet{Khan86}). In this model, the unit cell drawn in dashed lines in Fig.~\ref{Fig:Princen_T1_orientation} is sheared, and the location of the vertices is computed to comply with the equilibrium rule that the three edges meet at equal angles. To account for the finite liquid fraction, the vertices are decorated with Plateau borders which radius $R_P$ is an increasing function of the liquid fraction (Fig.~\ref{Fig:Princen_T1_orientation}, left): $\phi_l = (2\sqrt{3} - \pi)R_P^2/A_h$, with $A_h$ the area of one hexagon. This structure can be sheared up to the point where two neighbouring Plateau borders meet, which defines the onset of the T1 (Fig.~\ref{Fig:Princen_T1_orientation}, right).


The two angles $\theta'_a$ and $\theta'_d$ can be computed from simple geometry, when the two Plateau borders come into contact (Fig.~\ref{Fig:Princen_T1_orientation}, right). The length of the edge $c$ between the two bubbles about to detach is equal to $R_P$. Now at a given strain $\gamma$, this length equals \citep{Khan86}: $c = a(1 - \gamma\sqrt{3}/2)/\sqrt{4 + \gamma^2}$, where $a$ is the side length of the undeformed hexagon. Setting $c = R_P$ in the latter equation yields the strain $\gamma_c$ at which the T1 occurs; $\gamma_c$ is a decreasing function of $R_P$. Moreover, $\gamma = 1/\sqrt{3} - \cot\alpha$ \citep{Princen83}, and we compute from geometrical considerations in the right panel of Fig.~\ref{Fig:Princen_T1_orientation}: $\cot\theta'_d = 2/\sqrt{3} - \cot\alpha = 1/\sqrt{3} + \gamma_c$, and $\cot\theta'_a = -2/\sqrt{3} - \cot\alpha = -\sqrt{3} + \gamma_c$. Qualitatively, these two expressions show that both $\theta'_d$ and $\theta'_a$ are decreasing functions of $\gamma_c$, hence increasing functions of $R_P$, hence of the liquid fraction. The extreme values are found for a dry foam at $\phi_l = 0$, for which $\gamma_c = 2/\sqrt{3}$ \citep{Princen83}, and for the jamming transition for which $\gamma_c = 0$: $\theta_d'$ varies between $\pi/6 \simeq 0.52$~rad (dry foam) and $\pi/3 \simeq 1.05$~rad (jamming transition), and $\theta'_a$ between $-\pi/3$ and $-\pi/6$. Our measured values are indeed in these ranges. The values of the disappearing angles for the four experiments with $\phi_l = 4.8\%$ are compatible (within experimental dispersion) with the dry foam prediction, the latter indicated with a vertical dashed line in Fig.~\ref{Fig:T1_orientation}. The predicted increase of the angles with liquid fraction is compatible with the experiments for $\theta_d$, but not for $\theta_a$.

Although the model by \citet{Princen83} gives useful reference values, it is difficult to make a more quantitative comparison based on liquid fraction, because the distribution of liquid is specific to each system. In simulations, the films between droplets are thick, and contain a significant proportion of the liquid. In experiments, the distribution of water is complex because of the 3D structure of the bubbles; there is relatively more water close to the confining plates than in the midplane in between \citep{Surface1}. The hexagonal foam model of \citet{Princen83} is a good approximation of the structure of our experimental foams across the midplane between the top and bottom confining plates, but the liquid fraction across this plane, relevant in the hexagonal model, is significantly lower than the experimental liquid fraction. Moreover, the measurement of the appearing angle is less precise than that of the disappearing one, because the relaxation of the four bubbles after a T1 is fast; hence, the measurements made on the image after the topological rearrangement may not be representative of the configuration at the instant of a T1. This also explains why the dispersion is larger for $\theta_a$ than for $\theta_d$.

\section{Conclusions}\label{sec:conclusion}

We have reported on the first experimental study measuring the rate of plastic events in Poiseuille flows of foams. Experiments have been supplemented  by numerical simulations, capable to capture the realistic foam structure and to incorporate naturally the expected mesoscopic dynamics. We have addressed the relation between T1 distribution and macroscopic rheology and revealed a link between the localisation lengthscale of the velocity profiles and that of plastic events across the channel, confirming the relevance of cooperativity for foams \citep{Katgert10}. The use of numerical simulations allowed to study in a controlled way (something not easily feasible in the experiments) the effect of wall friction, helping to confirm its role in the emergence of an extra localisation for the velocity profiles, as predicted theoretically \citep{Barry11}. Our study highlighted that the elasticity gives rise to a complex near-to-wall dynamics which calls for focused studies both experimentally (in the spirit of the recent work by \citet{Mansard14}) and numerically, and for a more refined theoretical modelling of the boundary conditions. Finally, unprecedented results on the distribution of the orientation of plastic events show --- with good agreement between experiments and numerics --- that there is a non-trivial correlation with the underlying local shear strain; this suggests that more complex forms for the propagators invoked in theoretical models of soft-glassy materials \citep{Bocquet09} may be needed, with an explicit angular structure, especially in situation of non-homogeneous stress (as it is for Poiseuille flows).

The authors kindly acknowledge funding from the European Research Council under the EU Seventh Framework Programme (FP7/2007-2013) / ERC Grant Agreement no[279004]. We acknowledge computational support from CINECA (IT). MS and AS gratefully acknowledge M. Bernaschi for computational support and F. Bonaccorso for helpful visualisations of the flowing emulsions from the numerical simulations.

\begin{appendix}

\section{Pressure tensor in LBM simulations}\label{AppendixA}

In this Appendix we provide the technical details for the lattice Boltzmann pressure tensor used in equations (\ref{Gamma}) and (\ref{FILMTENSION}) to compute both the surface tension and the disjoining pressure at the non-ideal interface. Given the mechanical model for the lattice interactions described in (\ref{Phase})-(\ref{PSI}), an exact lattice theory is available \citep{Shan08,SbragagliaBelardinelli13} which allows to connect the interaction forces to the lattice Pressure Tensor. The exact pressure tensor is given by
$$
{P}_{\alpha \beta}=\sum_{\zeta, i} f_{\zeta i} {c}^{\alpha}_i {c}^{\beta}_i + \sum_{\zeta} {P}^{(int)}_{\zeta, \alpha \beta}.
$$
The term $\sum_{\zeta, i} f_{\zeta i} {c}^{\alpha}_i{c}^{\beta}_i$ represents an internal contribution to the pressure tensor, while ${P}^{(int)}_{\zeta, \alpha \beta}$ is a contribution coming from the interactions. We can separately write the contributions coming from the repulsive (r) phase separating interactions (see equation (\ref{Phase})), and those coming from competing interactions providing a mechanism of frustration (F) (see equation (\ref{NNandNNN}))
\be
{P}^{(int)}_{\zeta, \alpha \beta}={P}^{(r)}_{\zeta, \alpha \beta}+{P}^{(F,1)}_{\zeta, \alpha \beta}+{P}^{(F,2)}_{\zeta, \alpha \beta}+{P}^{(F,4)}_{\zeta, \alpha \beta}+{P^{(F,5)}_{\zeta, \alpha \beta}}+{P^{(F,8)}_{\zeta, \alpha \beta}}.
\ee
The contribution coming from the phase separating interactions ${P}^{(r)}_{\zeta, \alpha \beta}$ is \citep{SbragagliaBelardinelli13}
\be
{P}^{(r)}_{\zeta, \alpha \beta}=\frac{{\cal G}_{A B}}{2}\rho_{\zeta}(\mathbf{r})\sum_{i=1-8} w_i \rho_{\zeta'}(\mathbf{r}+\mathbf{c}_i){c}^{\alpha}_i{c}^{\beta}_i \hspace{.2in} \zeta' \neq \zeta
\ee
while the contributions coming from the frustrating interactions are given by various terms, ${P}^{(F,1)}_{\zeta,\alpha \beta}$, ${P}^{(F,2)}_{\zeta,\alpha \beta}$, ${P}^{(F,4)}_{\zeta,\alpha \beta}$, ${P^{(F,5)}_{\zeta,\alpha \beta}}$, ${P^{(F,8)}_{\zeta,\alpha \beta}}$, labeled with the number of the ``energy shell'' (see Tab.~\ref{TABLE})
\be
{P}^{(F,1)}_{\zeta, \alpha \beta}=\frac{{\cal G}_{\zeta \zeta, 1}}{2} \psi_{\zeta}(\mathbf{r}) \sum_{i=1-4} w_1 \psi_{\zeta}(\mathbf{r}+\mathbf{c}_i) {c}^{\alpha}_i {c}^{\beta}_i+\frac{{\cal G}_{\zeta \zeta, 2}}{2} \psi_{\zeta}(\mathbf{r}) \sum_{i=1-4} p_1 \psi_{\zeta}(\mathbf{r}+\mathbf{c}_i) {c}^{\alpha}_i {c}^{\beta}_i
\ee
\be
{P}^{(F,2)}_{\zeta, \alpha \beta}=\frac{{\cal G}_{\zeta \zeta, 1}}{2} \psi_{\zeta}(\mathbf{r}) \sum_{i=5-8} w_2 \psi_{\zeta}(\mathbf{r}+\mathbf{c}_i) {c}^{\alpha}_i {c}^{\beta}_i+\frac{{\cal G}_{\zeta \zeta, 2}}{2} \psi_{\zeta}(\mathbf{r}) \sum_{i=5-8} p_2 \psi_{\zeta}(\mathbf{r}+\mathbf{c}_i) {c}^{\alpha}_i {c}^{\beta}_i
\ee
\be
{P}^{(F,4)}_{\zeta, \alpha \beta}=\frac{{\cal G}_{\zeta \zeta, 2}}{4} \sum_{i=9-12} w_4  \psi_{\zeta}(\mathbf{r}) \psi_{\zeta}(\mathbf{r}+\mathbf{c}_{i}){c}_{i}^{\alpha} {c}_{i}^{\beta}+\frac{{\cal G}_{\zeta \zeta, 2}}{4} \sum_{i=9-12} w_4 \psi_{\zeta}\left(\mathbf{r}-\frac{\mathbf{c}_{i}}{2}\right)\psi_{\zeta}\left(\mathbf{r}+\frac{{c}_{i}}{2}\right)   {c}_{i}^{\alpha} {c}_{i}^{\beta}
\ee
\begin{equation}
\begin{split}
{P}^{(F,5)}_{\zeta, \alpha \beta}= \frac{{\cal G}_{\zeta \zeta, 2}}{4} \psi_{\zeta}(\mathbf{r}) \sum_{i=13-20} & p_5  \psi_{\zeta} (\mathbf{r}+\mathbf{c}_{i})  {c}_{i}^{\alpha} {c}_{i}^{\beta}+  \nonumber \\
+ \frac{{\cal G}_{\zeta \zeta, 2}}{4} & p_5 \left[\psi_{\zeta} (\mathbf{r}+\mathbf{c}_5) \psi_{\zeta} (\mathbf{r}+\mathbf{c}_3)+\psi_{\zeta} (\mathbf{r}+\mathbf{c}_1) \psi_{\zeta} (\mathbf{r}+\mathbf{c}_7) \right]{c}_{13}^{\alpha} {c}_{13}^{\beta} +\\
+ \frac{{\cal G}_{\zeta \zeta, 2}}{4} & p_5\left[ \psi_{\zeta} (\mathbf{r}+\mathbf{c}_5) \psi_{\zeta} (\mathbf{r}+\mathbf{c}_4)+\psi_{\zeta} (\mathbf{r}+\mathbf{c}_2) \psi_{\zeta} (\mathbf{r}+\mathbf{c}_7) \right]{c}_{14}^{\alpha} {c}_{14}^{\beta} + \nonumber \\
+ \frac{{\cal G}_{\zeta \zeta, 2}}{4} & p_5 \left[ \psi_{\zeta} (\mathbf{r}+\mathbf{c}_2) \psi_{\zeta} (\mathbf{r}+\mathbf{c}_8)+\psi_{\zeta} (\mathbf{r}+\mathbf{c}_6) \psi_{\zeta} (\mathbf{r}+\mathbf{c}_4) \right]{c}_{15}^{\alpha} {c}_{15}^{\beta}+      \nonumber \\
+ \frac{{\cal G}_{\zeta \zeta, 2}}{4} & p_5 \left[ \psi_{\zeta} (\mathbf{r}+\mathbf{c}_6) \psi_{\zeta} (\mathbf{r}+\mathbf{c}_1)+\psi_{\zeta} (\mathbf{r}+\mathbf{c}_3) \psi_{\zeta} (\mathbf{r}+\mathbf{c}_8) \right]{c}_{16}^{\alpha} {c}_{16}^{\beta}   \nonumber
\end{split}
\end{equation}
\be
{P}^{(F,8)}_{\zeta, \alpha \beta}=\frac{{\cal G}_{\zeta \zeta, 2}}{4} \psi_{\zeta}(\mathbf{r}) \sum_{i=21-24} p_8   \psi_{\zeta}(\mathbf{r}+\mathbf{c}_{i}){c}_{i}^{\alpha} {c}_{i}^{\beta}+\frac{{\cal G}_{\zeta \zeta, 2}}{4} \sum_{i=21-24} p_8 \psi_{\zeta}\left(\mathbf{r}-\frac{\mathbf{c}_{i}}{2}\right)\psi_{\zeta}\left(\mathbf{r}+\frac{\mathbf{c}_{i}}{2}\right)   {c}_{i}^{\alpha} {c}_{i}^{\beta}.
\ee



\begin{table}
\begin{center}
\begin{tabular}{|c|c|c|}
\hline
Phase Separating Interactions eq. (\ref{Phase}) & Shell & lattice Links\\
\hline
$ w_i  $& $|\mathbf{c}_{i}|^2$ &$ \mathbf{c}_{i} $ \\
\hline
$4/9$      & $0$ & $ (0,0)$\\
$1/9$      & $1$ & $ (\pm 1,0); (0,\pm 1)$\\
$1/36$     & $2$ & $  (\pm 1,\pm 1) $\\
\hline
\end{tabular}
\vspace{.2in}
\begin{tabular}{|c|c|c|c|}
\hline
1st term, rhs eq. (\ref{NNandNNN}) & 2nd term, rhs eq. (\ref{NNandNNN}) & Shell & lattice links\\
\hline
$w_i$ & $ p_i  $& $|\mathbf{c}_{i}|^2$ &$ \mathbf{c}_{i} $ \\
\hline
$4/9$  &  $247/420     $ & $0$ & $ (0,0)$\\
$1/9$  &  $4/63     $ & $1$ & $ (\pm 1,0); (0,\pm 1)$\\
$1/36$ & $4/135  $ & $2$ & $  (\pm 1,\pm 1) $\\
$0$ &  $1/180   $ & $4$ & $ (\pm 2,0); (0,\pm 2) $\\
$0$ &  $2/945   $ & $5$ & $  (\pm 2,\pm 1);(\pm 1,\pm 2)$\\
$0$ &  $1/15120  $ & $8$ & $ (\pm 2,\pm 2)$\\
\hline
\end{tabular}

\end{center}
\caption{\label{T1}\small{Links and weights of the two belts, $25$-speeds lattice \citep{Shan06,CHEM09} for all  interactions given in equations (\ref{Phase}) and (\ref{NNandNNN}). The first belt lattice velocities are indicated with $i=1...8$ while the second belt ones with $i=9...24$. $p_{i}$ or $w_i$ indicate the weight associated with the $i$-th link in the various interactions. The weights associated to the velocity at rest, $w_0$ and $p_0$, are chosen to enforce  a unitary normalisation, $\sum_{i=0}^{i=8} w_i=1$ and $\sum_{i=0}^{i=24} p_i=1$.} \label{TABLE}}
\end{table}

\section{Friction forces in LBM simulations}\label{AppendixB}

\begin{figure}
\begin{center}
\includegraphics[width=8.0cm,keepaspectratio]{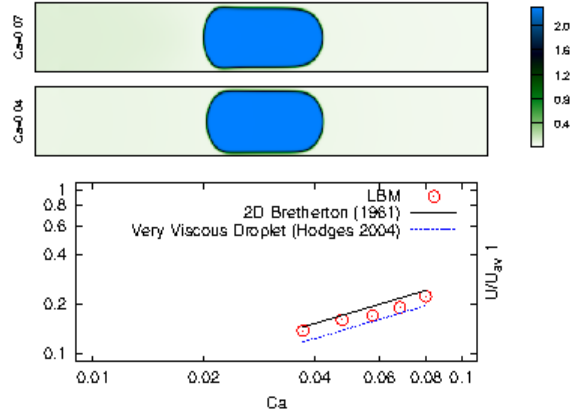}
\caption{Velocity of a 2D droplet in a confined channel. The viscous ratio between the dispersed phase and the continuous phase is set to $\chi=1$ in all the numerical simulations. In the top panel we report two snapshots  associated with two different capillary numbers. Blue/white (dark/light) colors indicate regions with majority of the dispersed/continuous phase. The droplet is driven by a constant pressure gradient. The average velocity of the droplet is normalised with respect to the mean flow velocity (${U}_{\mathrm{av}}$) in the inlet of the channel. The scaling laws for both very viscous ($\chi \gg 1$) droplet \citep{Schwartz,Hodges} and a bubble in a 2D channel are reported \citep{bretherton,Afkhami}. The velocity scaling well agrees with the theoretical prediction, confirming a scaling exponent in the capillary number close to $2/3$ and a numerical coefficient in between the two extreme cases ($\chi \gg 1$ and $\chi \ll 1$). \label{fig_bretherton}}
\end{center}
\end{figure}

In this Appendix, we propose benchmark tests for the LBM introduced in section \ref{sec:LBM} with regard to the viscous drag forces acting on individual bubbles. Friction properties in the thin films between neighbouring droplets/bubbles or between droplets/bubbles and the walls are important for both foams and concentrated emulsions \citep{Denkov06,Denkov08,Denkov09,Katgert09}. We start by presenting benchmark computations for the motion of droplets in confined channels at changing the capillary number. The drag force on a single bubble that slides past a solid wall was first investigated by \citet{bretherton} and has recently received renewed attention \citep{Denkov06,Denkov08,Katgert09}. For a single bubble sliding past a solid wall, \citet{bretherton} showed that the drag force scales non-linearly with the capillary number, $\mathrm{Ca}$, defined in terms of the dynamic viscosity of the carrier liquid and the relative velocity between the bubble and the wall. The lubrication approximation yields the velocity $U$ of the bubble immersed in the Poiseuille flow to be approximated at leading order in
\be\label{tube}
U/U_{\mathrm{av}}=1+1.29 (3 \mathrm{Ca})^{2/3} ,
\ee
where ${U}_{\mathrm{av}}$  represents the mean flow velocity. This theory can be readily modified for the 2D bubble placed in the Poiseuille flow between two parallel plates \citep{Afkhami}
\be\label{poiseuille}
U/{U}_{\mathrm{av}}=1+0.643 (3 \mathrm{Ca})^{2/3} .
\ee
An extension to droplets with an arbitrary viscosity has been considered in various papers \citep{Schwartz,Hodges}. For a very viscous drop, results analogous to the previous equations can readily be found, as the coefficients $1.29$ and $0.643$ in equations (\ref{tube}) and (\ref{poiseuille}), respectively, are reduced by a factor of $2^{-1/3}\approx 0.794$, yielding
\be\label{tubevisco}
U/{U}_{\mathrm{av}}=1+1.023 (3 \mathrm{Ca})^{2/3}
\ee
for a bubble in a cylindrical capillary and
\be\label{poiseuillevisco}
U/{U}_{\mathrm{av}}=1+0.511 (3 \mathrm{Ca})^{2/3}
\ee
for a 2D bubble in a channel. In Fig.~\ref{fig_bretherton} we present our benchmark tests for equation (\ref{poiseuillevisco}). As we can see, the predicted scaling for the velocity of the bubble well agrees with the theoretical prediction, confirming a scaling exponent in the capillary number close to $2/3$ and a numerical coefficient in between the case of a very viscous droplet and the case of a bubble

\begin{figure}
\begin{center}
\includegraphics[scale=0.6]{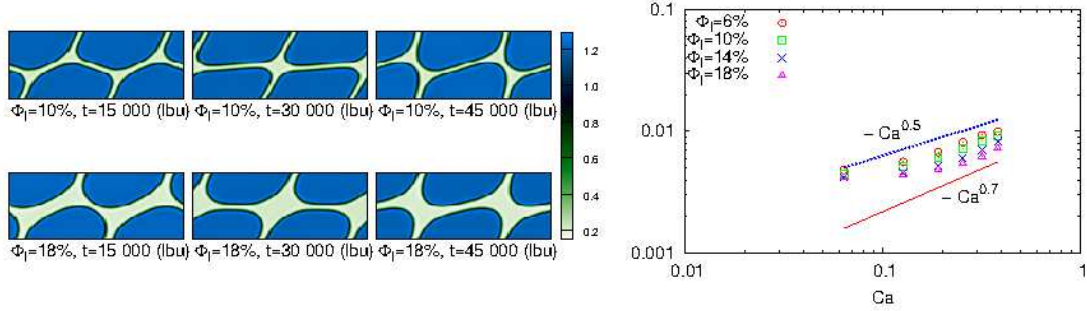}
\caption{Viscous drag force $F_{bb}$ between droplets measured directly in rheological experiments where two rows of ordered bubbles are sheared past each other. The packing fraction of the continuous phase into the dispersed phase is changed in the interval $\phi_{l}=[0.06:0.18]$. The left panel reports three snapshots of the simulations for two different packing fractions.  Blue/white (dark/light) colors indicate regions with majority of the dispersed/continuous phase. A scaling law in the Capillary number is found, $F_{bb} \sim \mathrm{Ca}^{\xi}$, with a scaling exponent between $1/2$ and $2/3$. \label{fig_drag}}
\end{center}
\end{figure}

We now continue by presenting benchmark tests for the drag force between two bubbles sliding past each other, $F_{bb}$. Some recent works \citep{Denkov06,Denkov08,Katgert09} have provided evidences that the viscous drag force scales like $F_{bb} \propto \mathrm{Ca}^{\xi}$, with a scaling exponent $\xi$ between $1/2$ and $2/3$. This is an important test for our numerical simulations: LBM modelling of two phase flows is intrinsically a diffuse interface method and involves a finite thickness of the interface between the two liquids and related model parameters. The values of the interface thickness and capillary number need to be larger than the one suggested by physical considerations in order to make the simulations affordable \citep{Komrakova,Casciola}. Nevertheless, the structure and the dynamical properties of the emulsion droplets that we reproduce in the numerical simulations share nontrivial features with the experiments \citep{Goyon08,Goyon10,Mansard14}. It is therefore of great importance  to investigate the scaling laws associated with friction properties, to show that they are realistic and in line with those measured in experiments. In particular, we measure the viscous drag forces between bubbles directly by rheological experiments where two rows of ordered bubbles are sheared past each other. Results are reported in Fig.~\ref{fig_drag}. A scaling law in the velocity difference $U$ between the two rows of droplets is confirmed, $F_{bb} \sim \mathrm{Ca}^{\xi}$, with a scaling exponent between $1/2$ and $2/3$.


\section{Velocity profile in the nonlinear local model} \label{AppendixC}

In this Appendix, we provide a solution of the velocity profile, obeying momentum conservation (\ref{Eq:general_equation_for_stress}) with the nonlinear law (\ref{Eq:nonlinear_friction}) for friction, and (\ref{Eq:nonlinear_viscous_stress}) for the viscous stress but with a velocity exponent $1/2$ instead of 0.47, an approximation which enables to provide an analytical solution. Hence we solve for $-H/2 \leq y\leq 0$ (so that $\D v/\D y \geq 0$):
$$ 0 = \eta' \frac{\D}{\D y} \left[ \left( \frac{\D v}{\D y} \right)^{1/2} \right] - \frac{\D P}{\D x} - \beta' v^{1/2} , $$
with boundary conditions: $\D v/\D y = 0$ at $y = 0$, and $\sigma = -f_v$ at $y = -H/2$.

We make space and velocity dimensionless: $\bar{y} = y/Y$ and $\bar{v} = v/V$, by rescaling them by
\be\label{eq:rescaling}
Y = (\eta'/\beta')^{2/3} \hspace{.3in} V = [(-\D P/\D y)/\beta]^2.
\ee
Then the problem becomes:
\begin{equation}\label{Eq:dimensionless_ODE_velocity}
0 = (\bar{v}_{\bar{y}}^{1/2})_{\bar{y}} - \bar{v}^{1/2} + 1 ,
\end{equation}
where the subscript designs derivation, with boundary conditions: $\bar{v}_{\bar{y}} = 0$ at $\bar{y} = 0$, and $\bar{v}_{\bar{y}}^{1/2} = h\bar{v}^{1/2}/2Y$ at $\bar{y} = -H/2Y$. The velocity thus obeys an autonomous equation of the form: $\bar{v}_{\bar{y}\bar{y}} = F(\bar{v},\bar{v}_{\bar{y}})$, with $F(x,y) = -2(1 - \sqrt{x})\sqrt{y}$. This kind of equation can be recast as a first-order ODE \citep{Polyanin03} by setting $\bar{w} = \bar{v}_{\bar{y}}$: it then becomes $\bar{w}_{\bar{v}} = F(\bar{v},\bar{w})/\bar{w} = -2(1 - \sqrt{\bar{v}})/\sqrt{\bar{w}}$. The latter is an ODE with separable variables, which is thus simply integrated to yield: $2\bar{w}^{3/2}/3 = -2\bar{v} + 4\bar{v}^{3/2}/3 + \mathrm{const}$. The boundary condition at $\bar{y} = 0$ imposes that $\bar{w} = 0$ for the unknown centreline velocity $\bar{v}_0$, hence $2\bar{w}^{3/2}/3 = -2(\bar{v} - \bar{v}_0) + 4(\bar{v}^{3/2} - \bar{v}_0^{3/2})/3$. In the limit $2Y/H \ll 1$, which is a good approximation in our experiments, the dimensionless viscous stress term $(\bar{v}_{\bar{y}}^{1/2})_{\bar{y}}$ is negligible at the centre of the channel, hence after (\ref{Eq:dimensionless_ODE_velocity}), $\bar{v}_0 = 1$. Therefore,
\begin{equation}\label{Eq:dimensionless_ODE_velocity_order_1}
\bar{v}_{\bar{y}} = [(1 - \sqrt{\bar{v}})^2 (2\sqrt{\bar{v}} + 1)]^{2/3} .
\end{equation}
The r.h.s. is a decreasing function over $[0,1]$, equal to 1 for $\bar{v} = 0$ and to 0 for $\bar{v} = 1$. Therefore, for a given value of the parameter $h/2Y$, the boundary condition at the wall: $\bar{v}_{\bar{y}} = h^2 \bar{v}/4Y^2$, admits a single solution for $\bar{v}$ and $\bar{v}_{\bar{y}}$ at $\bar{y} = -H/2Y$. The velocity field obeying (\ref{Eq:dimensionless_ODE_velocity}) and the boundary condition then obeys the implicit equation: $\bar{y} + H/2Y = \Phi(\bar{v}) - \Phi(\bar{v}_s)$, with:
\begin{eqnarray*}
  \Phi(\bar{v}) &=& \int \frac{\D\bar{v}}{[(1 - \sqrt{\bar{v}})^2 (2\sqrt{\bar{v}} + 1)]^{2/3}} \\
   &=& (1 + 2\sqrt{\bar{v}})^{1/3} \left[ \frac{2}{(1 - \sqrt{\bar{v}})^{1/3}} + 2^{1/3} 3^{2/3} {_2 F_1} \left( \frac{1}{3},-\frac{2}{3},\frac{4}{3},\frac{1}{3}(1 + 2\sqrt{\bar{v}}) \right) \right. \\ && \qquad \left. - \frac{2^{1/3} 3^{2/3}}{12} {_2 F_1} \left( \frac{4}{3},\frac{1}{3},\frac{7}{3},\frac{1}{3}(1 + 2\sqrt{\bar{v}}) \right) \right] ,
\end{eqnarray*}
where ${_2 F_1}$ designs the hypergeometric function. With this complicated expression, fitting the experimental data is not easy. A simpler alternative consists in developing (\ref{Eq:dimensionless_ODE_velocity_order_1}) for $1 - \bar{v} \ll 1$. This gives: $\bar{v}_{\bar{y}} = (3/4)^{2/3} (1 - \bar{v})^{4/3}$, with general solution: $1 - \bar{v} = 48/(\bar{y} + \mathrm{const}.)^3$. This gives an alternative fitting formula for the velocity profile:
\begin{equation}\label{Eq:fit_nonlinear_laws}
v = V \left[ 1 - \frac{48Y^3}{(y - y_0)^3} \right] .
\end{equation}
Taking $V$, $Y$ and $y_0$ yields a fit which is close to the exponential fit (\ref{Eq:fit_velocity}), the difference between both fits being within the dispersion of the experimental data. This suggests that the effect of the nonlinearities of the friction and of the viscous stress on the flow profile is weak. Finally, we have checked that the qualitative conclusions brought by Fig.~\ref{Fig:Slip_velocity_exp_Janiaud}, namely that the relative slip is overestimated by the local model, remains valid with nonlinear laws. Hence, the role of non-linearities is secondary in this study, and can be neglected.

\end{appendix}

\bibliographystyle{jfm}

\end{document}